\documentclass[twocolumn]{aastex631}

\usepackage [english]{babel}
\usepackage [autostyle, english = american]{csquotes}
\MakeOuterQuote{"}
\usepackage[figuresright]{rotating}
\usepackage{longtable}
\usepackage{makecell}
\usepackage{graphicx}
\usepackage{rotating}

\begin{document}

\title{Near-Infrared Flux Distribution of Sgr A* from 2005-2022: Evidence for an Enhanced Accretion Episode in 2019}

\correspondingauthor{Grant C. Weldon}
\email{gweldon@astro.ucla.edu}

\author{Grant C. Weldon}
\affiliation{UCLA Galactic Center Group, Physics and Astronomy Department, University of California, Los Angeles, CA 90095}

\author{Tuan Do}
\affiliation{UCLA Galactic Center Group, Physics and Astronomy Department, University of California, Los Angeles, CA 90095}

\author{Gunther Witzel}
\affiliation{Max Planck Institute for Radio Astronomy, Auf dem Hügel 69, D-53121 Bonn (Endenich), Germany}

\author{Andrea M. Ghez}
\affiliation{UCLA Galactic Center Group, Physics and Astronomy Department, University of California, Los Angeles, CA 90095}

\author{Abhimat K. Gautam}
\affiliation{UCLA Galactic Center Group, Physics and Astronomy Department, University of California, Los Angeles, CA 90095}

\author{Eric E. Becklin}
\affiliation{UCLA Galactic Center Group, Physics and Astronomy Department, University of California, Los Angeles, CA 90095}

\author{Mark R. Morris}
\affiliation{UCLA Galactic Center Group, Physics and Astronomy Department, University of California, Los Angeles, CA 90095}

\author{Gregory D. Martinez}
\affiliation{UCLA Galactic Center Group, Physics and Astronomy Department, University of California, Los Angeles, CA 90095}

\author{Shoko Sakai}
\affiliation{UCLA Galactic Center Group, Physics and Astronomy Department, University of California, Los Angeles, CA 90095}

\author{Jessica R. Lu}
\affiliation{University of California, Berkeley, Department of Astronomy, Berkeley, CA 94720}

\author{Keith Matthews}
\affiliation{Division of Physics, Mathematics, and Astronomy, California Institute of Technology,
MC 301-17, Pasadena, California 91125, USA}

\author{Matthew W. Hosek Jr.}
\affiliation{UCLA Galactic Center Group, Physics and Astronomy Department, University of California, Los Angeles, CA 90095}

\author{Zoë Haggard}
\affiliation{UCLA Galactic Center Group, Physics and Astronomy Department, University of California, Los Angeles, CA 90095}

\begin{abstract}
Sgr A* is the variable electromagnetic source associated with accretion onto the Galactic center supermassive black hole. While the near-infrared (NIR) variability of Sgr A* was shown to be consistent over two decades, unprecedented activity in 2019 challenges existing statistical models. We investigate the origin of this activity by re-calibrating and re-analyzing all of our Keck Observatory Sgr A* imaging observations from 2005-2022. We present light curves from 69 observation epochs using the NIRC2 imager at 2.12 $\mu$m with laser guide star adaptive optics. These observations reveal that the mean luminosity of Sgr A* increased by a factor of $\sim$3 in 2019, and the 2019 light curves had higher variance than in all time periods we examined. We find that the 2020-2022 flux distribution is statistically consistent with the historical sample and model predictions, but with fewer bright measurements above 0.6 mJy at the $\sim$2$\sigma$ level. Since 2019, we have observed a maximum $K_s$ (2.2 $\mu$m) flux of 0.9 mJy, compared to the highest pre-2019 flux of 2.0 mJy and highest 2019 flux of 5.6 mJy. Our results suggest that the 2019 activity was caused by a temporary accretion increase onto Sgr A*, possibly due to delayed accretion of tidally-stripped gas from the gaseous object G2 in 2014. We also examine faint Sgr A* fluxes over a long time baseline to search for a quasi-steady quiescent state. We find that Sgr A* displays flux variations over a factor of $\sim$500, with no evidence for a quiescent state in the NIR.
\end{abstract}

\keywords{black hole physics, accretion, Galaxy: center, techniques: high angular resolution}

\section{Introduction} \label{sec:intro}

The Milky Way's Galactic center harbors the nearest supermassive black hole, providing a natural laboratory for detailed study of physical processes in its vicinity \citep[e.g.][]{Schodel03, Ghez08}. Observations at radio, infrared, and X-ray wavelengths have revealed the Galactic black hole, Sgr A*, to be a continuously variable source \citep[e.g.][]{Falcke99, Baganoff01, Genzel03}. The detailed nature of the variable emission can provide important constraints on how gas is captured from the black hole’s surrounding environment, as well as on the physical phenomena taking place within the accretion flow as the matter approaches the event horizon (e.g., reconnection events, hydrodynamic and magnetohydrodynamic instabilities, shocks, and lensing caustics). While accretion processes have been extensively studied in highly luminous active galactic nuclei, Sgr A* is an example of the ubiquitous population of low-luminosity galactic nuclei in the nearby universe. By studying Sgr A*, we can learn more about accretion processes in these systems \citep[for recent reviews see][]{Genzel10, Morris12}.

Near-infrared (NIR) observations are particularly effective for monitoring the variability of Sgr A*. While the radio band traces small variations in the bulk accretion flow and the X-ray band is only sensitive to strong flares that rise above the extended local background at the Galactic center, the resolved NIR source shows both large and small variations that can almost always be detected using adaptive optics on the largest ground-based telescopes \citep[e.g.][]{Hornstein02, Genzel03, Ghez04, Meyer14}. The NIR emission arises primarily from synchrotron radiation very close ($\lesssim$15 Schwarzschild radii) to the event horizon and can be used to bridge the radio and X-ray bands to construct a coherent description of the emission processes in radiatively-inefficient, advection-dominated black hole accretion flows \citep[e.g.][]{Fazio18, Witzel21}. 

There have been numerous efforts to statistically characterize the NIR variability. \cite{Do09} showed that light curves from 2005-2007 can be described by a red-noise process that is correlated in time and does not exhibit periodicity or quasi-periodicity. Using observations from 2004-2009, \cite{Dodds-Eden11} found the data to be described by a two-state model with a low-level log-normal component and a separate power-law tail for flares. On the other hand, a recent comprehensive analysis of over 13,000 measurements of Sgr A* flux density with Keck data at 2.12 $\mu$m (2003-2014) and VLT and Spitzer data at 4.5 $\mu$m (2004-2017) found that the NIR time variability can indeed be consistently described by red noise, but that the entire distribution of fluxes is well-fit by a single log-normal model without a separate flaring component \citep{Witzel18}. By re-analyzing speckle data from Keck (1995-2005), \cite{Chen19} extended the time baseline of Sgr A* observations and showed that the statistical model of \cite{Witzel18} remained consistent with the data over two decades through 2018.

However, in May 2019 \cite{Do19} observed an unprecedented bright NIR event with Keck. The flux of Sgr A* varied by a factor of 75 over a 2-hour timescale and reached twice the level of any previous historical NIR measurement. The maximum fluxes were observed at the beginning of observations, suggesting that Sgr A* was likely even brighter prior to taking measurements that night. With additional bright events observed at Keck in 2019, \cite{Do19} showed the 2019 activity to be inconsistent with the historical flux distribution of \cite{Witzel18}. Additionally, a high abundance of bright NIR events were observed with VLT in 2019 relative to other years \citep{GRAVITY20}. Heightened activity was also seen in the radio \citep[e.g.][]{Murchikova21b, Boyce22, Cheng23} and X-ray regimes \citep[e.g.][]{Pavlinsky19, Degennar19}. These observations collectively suggest that Sgr A* displayed unusual behavior in 2019, making this era useful for understanding the time-variable accretion flow onto the black hole. 

Various hypotheses have been proposed to explain the elevated 2019 activity. These possibilities and their observable signatures include the following. 

\begin{enumerate}
\item The statistical models of Sgr A* in the NIR need to be updated to capture the bright events. While \cite{Witzel18} found that historical measurements follow a single log-normal distribution, \cite{GRAVITY20} argues that the 2019 bright events observed by Keck and VLT are rare events drawn from the power-law tail of the underlying two-state distribution described by \cite{Dodds-Eden11}. In such two-state models, the faint distribution arises from steady accretion of mass from the black hole's environment, and the flaring state arises from a separate process that injects additional energy. A related question concerns whether there exists a quasi-steady, quiescent "floor" to the Sgr A* light curves or whether Sgr A* displays similar variability characteristics at both faint and high levels \citep[e.g.][]{Do09}. Previous studies have historically been complicated by short time baselines or have not closely examined the faint end of the flux distribution. We can examine Sgr A* at low flux densities over a long time baseline to determine whether such a quiescent state exists and to study whether the statistical character of the emission at faint flux densities is simply an extension of what we see at higher flux densities. We can also study whether the faint flux densities are different in 2019 than in other years. If the 2019 events are indicative of a power-law "flaring" state distinct from the log-normal "quiescent" state, we should see only the flaring state affected and the faint end of the flux distribution unchanged between 2019 and other years. A shift in the faint end of the flux distribution in 2019 would instead suggest that Sgr A* experienced a physical change during this time.

\item There was a temporary accretion increase onto Sgr A*. The Galactic center hosts a population of stars and dusty G-objects, which could affect the accretion flow. The star S0-2 underwent closest approach in 2018, and its passage may have induced an accretion disturbance \citep{Loeb04}. In the past two decades, the objects G1 and G2 underwent closest approach and showed signs of tidal interaction with the black hole, causing much speculation about the potential for excess material from these objects to induce heightened Sgr A* activity \citep[e.g.][]{Gillessen12, Phifer13, Eckart13, Witzel14, Pfhul15, Witzel17, Plewa17, Gillessen19}. While an increase in X-ray flares was reported following G2's passage in 2014 \citep{Ponti15}, no exceptional NIR activity was reported in the years immediately following 2014. However, it is possible that there is a larger time delay between the passage of G2 and any observed NIR activity \citep[e.g][]{Kawashima17, Murchikova21}. If a neighboring source, such as S0-2 or G2, deposited gas or altered the accretion flow onto Sgr A*, we would expect to see heightened activity for the duration of time it takes that material to fall into the black hole, followed by a return to more typical behavior. 

\item Sgr A* has entered a new accretion state in which bright flares are more common, or the 2019 activity is a precursor to an even larger event \citep{Do19}. Observations of X-ray light echoes from iron line emission at the Galactic center suggest that in the past several hundred years, Sgr A* has undergone multiple ($\sim$2-10 year) luminosity increases by factors of up to $10^5$ \citep[e.g.][]{Ponti10, Clavel13, Terrier18, Chuard18}. If Sgr A* is undergoing a state change or entering an extreme episode, we should see the activity of Sgr A* continue or evolve in the years beyond 2019.
\end{enumerate}

To test these hypotheses, we must compare the behavior of Sgr A* before, during, and after 2019 at both low and high flux densities. Previous studies used different photometric methods (e.g. aperture photometry, point spread function fitting, interferometry) and varying treatments for stellar contamination of Sgr A*, making the comparison of absolute flux densities unreliable, especially at fainter levels. In this work, we present Sgr A* observations using Keck Observatory from 2005-2022, all reduced with the same methodology, allowing for a robust comparison of flux densities over time. This study includes re-reduced data from \cite{Witzel18} and \cite{Do19}, as well as newly presented historical data (2005-2017; Sgr A* is confused by the bright star S0-2 in 2018) and more recent observations taken in 2019 and from 2020-2022. Section \ref{sec:obs} describes these observations and data reduction. Section \ref{sec:analysis} presents the analysis of the flux distributions, comparisons to the historical model, characteristics of the light curves, and timing analyses. Section \ref{sec:physics} discusses the physical implications of our results. We conclude in Section \ref{sec:conclusion}.

\section{Observations and Data Reduction} \label{sec:obs}

\subsection{Keck observations}

Our observations of the Galactic center were taken with the Keck 2 Telescope using the narrow camera in the Near-Infrared Camera 2 (NIRC2) imager with the Laser-Guide Star Adaptive Optics (LGS AO) system \citep{Wizinowich06}. All of the Keck observations in this work were taken with the $K'$ filter (2.12 $\mu$m). Individual $K'$ images consist of 10 coadds of 2.8-second integration time each. Standard image reduction methods were applied, including flat-fielding, sky subtraction, and cosmic ray removal via the Keck AO Imaging (KAI) data reduction pipeline \citep{KAI}.

Photometry was performed on individual images to construct light curves for Sgr A*. We used the point spread function (PSF) fitting program \textit{Starfinder} to measure the brightness and position of sources \citep{Diolaiti00}. Using the same procedure as in \cite{Do19}, a different PSF was constructed for each image to account for changes in effective PSF due to seeing variations. To increase the number of Sgr A* detections at faint flux levels, we used an enhanced version of \textit{Starfinder} that includes \textit{a priori} knowledge of the location of Sgr A* and other nearby NIR sources \citep{Hornstein07}. For each epoch, the source locations in a nightly-averaged image were used as fixed inputs into \textit{Starfinder} to more accurately fit for Sgr A* and neighboring sources in the individual frames.

Our photometric calibration has been updated since \cite{Do19}, making the observed flux densities of all sources within the central arcsecond of the Galactic center (including Sgr A*) systematically lower by $\sim$10\%. Photometric uncertainties were estimated by fitting a power law between the flux and RMS flux uncertainty for stars within 1 arcsecond of Sgr A*, as described in \cite{Do09}. We then used this relationship to calculate the uncertainty in flux for Sgr A*. More details about our photometric calibration and uncertainty estimations can be found in Appendix \ref{app:reduction}. The photometric stability of our observations is discussed in Appendix \ref{app:stability}.

In this study, we work with observed flux densities (rather than dereddened) to make our results comparable to \cite{Do19}. We convert our observed $K'$ fluxes to observed $K_s$ fluxes using the filter transformation $F_{K_s} = 1.09 F_{K'}$ \citep{Do19}. Dereddened fluxes can be computed using $F_{\text{dereddened}} = F_{\text{obs}} \times 10^{0.4 A_{K_{s}}}$. Different studies have used different extinction values, such as $A_{K_s} = 2.46$ \citep{Schodel10}, $A_{K_s} = 2.8$ \citep{Genzel03, Eckart06}, $A_{K_s} = 3.2$ \citep{Hornstein07}, and $A_{K_s} = 3.3$ \citep{Do09}. As in \cite{Do19}, we will occasionally aid readers by presenting both observed and dereddened flux densities computed using $A_{K_s} = 2.46$.

\subsection{Sample selection} \label{subsec:sample}

We introduce several selection criteria to make our Keck data set more robust for comparing Sgr A* variability over time. Beginning with 104 Galactic center observation epochs from 2005-2022, we apply the following successive cuts to our sample:
\begin{enumerate}
    \item Considering only observing epochs in which 20 or more quality (Strehl ratio $>$0.2) images of the Galactic center were taken. Below 20 frames, the light curves are poorly sampled for timing analyses. 18 such epochs are removed from our sample.

    \item Removing epochs in which our PSF reference stars are saturated. Saturated PSF reference stars introduce a photometric bias that makes our Sgr A* flux measurements unreliable. 3 such epochs are removed from our sample.

    \item Removing epochs in which the bright star S0-2 ($K$ = 14.2 mag) is confused with Sgr A* (in late 2017 and 2018). Attempting to correct for S0-2's flux bias introduces large uncertainties that are comparable to faint Sgr A* flux levels. 14 such epochs are removed, leaving 69 observation epochs for consideration. For epochs in which a fainter S-star is confused with Sgr A*, we apply a photometric correction (see Appendix \ref{app:confusion} for procedure).

    \item Considering only images with Strehl ratio greater than 0.2 within each of the 69 remaining nights. At lower Strehl ratios, the photometry is unreliable due to poor seeing conditions. About 10\% of 7935 frames have Strehl ratios below 0.2, leaving 7155 frames that survive our data quality cut. 
\end{enumerate}

The 69 observation epochs that we have selected for our sample are described in Table \ref{tab:observations}. The epochs that we have chosen to exclude and justifications for omission are found in Appendix \ref{app:omissions}. The light curve data are presented in Appendix \ref{app:data}.

\onecolumngrid
\begin{longtable*}{c c c c c c c c}
\caption{Summary of Sgr A* observations}
\label{tab:observations}
\tabularnewline 
\hline Date & $N_{\text{frames}}$ & $N_{\text{detections}}$ & Duration & Med. $F_{\text{obs}}$ & Max. $F_{\text{obs}}$ & Avg. Unc. & Med. Strehl\\ (UT) & & & (min) & (mJy) & (mJy) & (\%) & \\
\hline
\hline
\endfirsthead
\caption*{Summary of Sgr A* observations (continued)}
\tabularnewline 
\hline Date & $N_{\text{frames}}$ & $N_{\text{detections}}$ & Duration & Med. $F_{\text{obs}}$ & Max. $F_{\text{obs}}$ & Avg. Unc. & Med. Strehl\\ (UT) & & & (min) & (mJy) & (mJy) & (\%) & \\
\hline
\hline
\endhead
2005-07-31 & 28  & 28  & 119 & 0.09 & 0.41 & 11 & 0.39 \\ 
2006-05-03 & 121 & 121  & 135 & 0.10 & 0.65 & 3 & 0.37\\
2006-06-20 & 90  & 88  & 127 & 0.06 & 0.59 & 13 & 0.28\\
2006-06-21 & 157 & 157  & 164 & 0.03 & 0.27 & 10 & 0.38\\
2006-07-17 & 70  & 70  & 189 & 0.00 & 0.18 & 8 & 0.39\\
2007-05-17 & 87  & 87  & 164 & 0.04 & 0.43 & 10 & 0.38\\
2007-08-10 & 43  & 43  & 88  & 0.21 & 0.52 & 5 & 0.25\\
2007-08-12 & 58  & 58  & 55  & 0.09 & 0.17 & 2 & 0.36\\
2008-05-15 & 137 & 137 & 153 & 0.13 & 0.22 & 5 & 0.31\\
2008-07-24 & 167 & 167  & 177 & 0.02 & 0.21 & 10 & 0.32\\
2009-05-01 & 195 & 195 & 186 & 0.21 & 0.45 & 5 & 0.32\\
2009-05-02 & 54  & 54  & 61  & 0.30 & 0.51 & 4 & 0.39\\
2009-05-04 & 57  & 57  & 52  & 0.38 & 0.70 & 5 & 0.45\\
2009-07-22 & 22  & 22  & 27  & 0.28 & 0.41 & 8 & 0.23\\
2009-07-24 & 119 & 119 & 135 & 0.15 & 0.24 & 9 & 0.27\\
2009-09-09 & 49  & 49  & 56  & 0.06 & 0.19 & 18 & 0.38\\
2010-05-04 & 115 & 115 & 184 & 0.17 & 0.60 & 11 & 0.33\\
2010-05-05 & 73  & 73  & 67  & 0.54 & 2.01 & 4 & 0.35\\
2010-07-06 & 135 & 135 & 126 & 0.14 & 0.67 & 8 & 0.33\\
2010-08-15 & 142 & 72  & 135 & 0.04 & 0.14 & 13 & 0.32\\
2011-05-27 & 144 & 143 & 159 & 0.18 & 0.48 & 9 & 0.29\\
2011-07-18 & 202 & 200 & 204 & 0.10 & 0.33 & 12 & 0.28 \\
2011-08-23 & 108 & 107  & 93  & 0.06 & 0.18 & 15 & 0.38\\
2011-08-24 & 110 & 110 & 97  & 0.13 & 0.32 & 8  & 0.33\\
2012-05-15 & 207 & 83  & 184 & 0.05 & 0.22 & 14 & 0.33\\
2012-05-18 & 79  & 42  & 144 & 0.04 & 0.09 & 18 & 0.27\\
2012-07-24 & 210 & 203 & 201 & 0.14 & 0.88 & 10 & 0.24\\
2013-04-26 & 63  & 56  & 65  & 0.08 & 0.26 & 17 & 0.23\\
2013-04-27 & 77  & 57  & 130 & 0.05 & 0.24 & 25 & 0.23\\
2013-07-20 & 231 & 229 & 208 & 0.07 & 0.13 & 14 & 0.26\\
2014-03-20 & 20  & 20  & 45  & 0.08 & 0.12 & 11 & 0.24\\
2014-05-19 & 165 & 165 & 154 & 0.11 & 0.23 & 11 & 0.26\\
2014-07-03 & 46  & 46  & 140 & 0.09 & 0.24 & 14 & 0.23\\
2014-07-04 & 78  & 78  & 121 & 0.10 & 0.19 & 12 & 0.24\\
2014-08-04 & 26  & 26  & 62  & 0.15 & 0.25 & 10 & 0.24\\
2014-08-06 & 135 & 135 & 119 & 0.13 & 1.02 & 9 & 0.29\\
2015-03-31 & 43  & 43  & 124 & 0.13 & 0.35 & 9 & 0.24\\
2015-04-02 & 20  & 19  & 113 & 0.06 & 0.16 & 12 & 0.32\\
2015-08-09 & 99  & 99  & 120 & 0.33 & 0.94 & 8 & 0.31\\
2015-08-10 & 110 & 100  & 105 & 0.07 & 1.66 & 6 & 0.40\\
2015-08-11 & 89  & 89  & 77  & 0.07 & 0.13 & 12 & 0.41\\
2016-05-03 & 199 & 183 & 186 & 0.07 & 0.24 & 13 & 0.24\\
2016-07-13 & 186 & 168 & 190 & 0.08 & 0.21 & 14 & 0.25\\
2017-05-04 & 148 & 58  & 199 & 0.03 & 0.11 & 23 & 0.27\\
2017-05-05 & 254 & 202 & 250 & 0.13 & 0.78 & 8 & 0.36\\
2019-04-19 & 60  & 46  & 101 & 0.12  & 0.35 & 15 & 0.23\\
2019-04-20 & 168 & 168 & 149 & 0.26 & 1.56 & 7 & 0.34\\
2019-05-13 & 90  & 87  & 213 & 0.18 & 5.58 & 8 & 0.35\\
2019-05-23 & 168 & 168 & 177 & 0.19 & 0.64 & 9 & 0.35\\
2019-08-14 & 39  & 31  & 124 & 0.13 & 0.53 & 9 & 0.29\\
2019-08-19 & 24  & 23  & 20  & 0.10 & 0.29 & 11 & 0.29\\
2020-07-07 & 197 & 170  & 237 & 0.05 & 0.18 & 24 & 0.28\\
2020-08-09 & 32  & 32  & 88  & 0.34 & 0.55 & 10 & 0.26\\
2021-05-13 & 52  & 52  & 90  & 0.14 & 0.25 & 4  & 0.23\\
2021-05-14 & 146 & 146 & 152 & 0.19 & 0.54 & 3  & 0.31\\
2021-07-13 & 32  & 32  & 117 & 0.29 & 0.45 & 5  & 0.23\\
2021-07-14 & 35  & 35  & 216 & 0.24 & 0.60 & 7  & 0.22\\
2021-08-13 & 136 & 136 & 141 & 0.16 & 0.22 & 5  & 0.38\\
2022-05-14 & 175 & 116 & 201 & 0.07 & 0.33 & 10 & 0.37\\
2022-05-15 & 27  & 27  & 174 & 0.22 & 0.74 & 10 & 0.24\\
2022-05-21 & 51  & 27  & 105 & 0.05 & 0.93 & 7 & 0.44\\
2022-05-25 & 65  & 44  & 132 & 0.08 & 0.15 & 9 & 0.38\\
2022-07-19 & 133 & 127 & 141 & 0.11 & 0.28 & 5 & 0.33\\
2022-07-22 & 227 & 114 & 219 & 0.05 & 0.19 & 8 & 0.39\\
2022-08-14 & 66  & 52  & 64  & 0.11 & 0.22 & 8 & 0.29\\
2022-08-15 & 124 & 77  & 110 & 0.06 & 0.32 & 4 & 0.48\\
2022-08-16 & 39  & 39  & 99  & 0.16 & 0.27 & 9 & 0.31\\
2022-08-19 & 49  & 49  & 114 & 0.25 & 0.48 & 4 & 0.41\\
2022-08-20 & 52  & 52  & 116 & 0.22 & 0.50 & 7 & 0.31\\
\hline
\hline
\end{longtable*}
\twocolumngrid

\section{Results and Analysis} \label{sec:analysis}

\subsection{Significantly different flux distribution in 2019}

\begin{figure}[ht]

\includegraphics[width=3.3in]{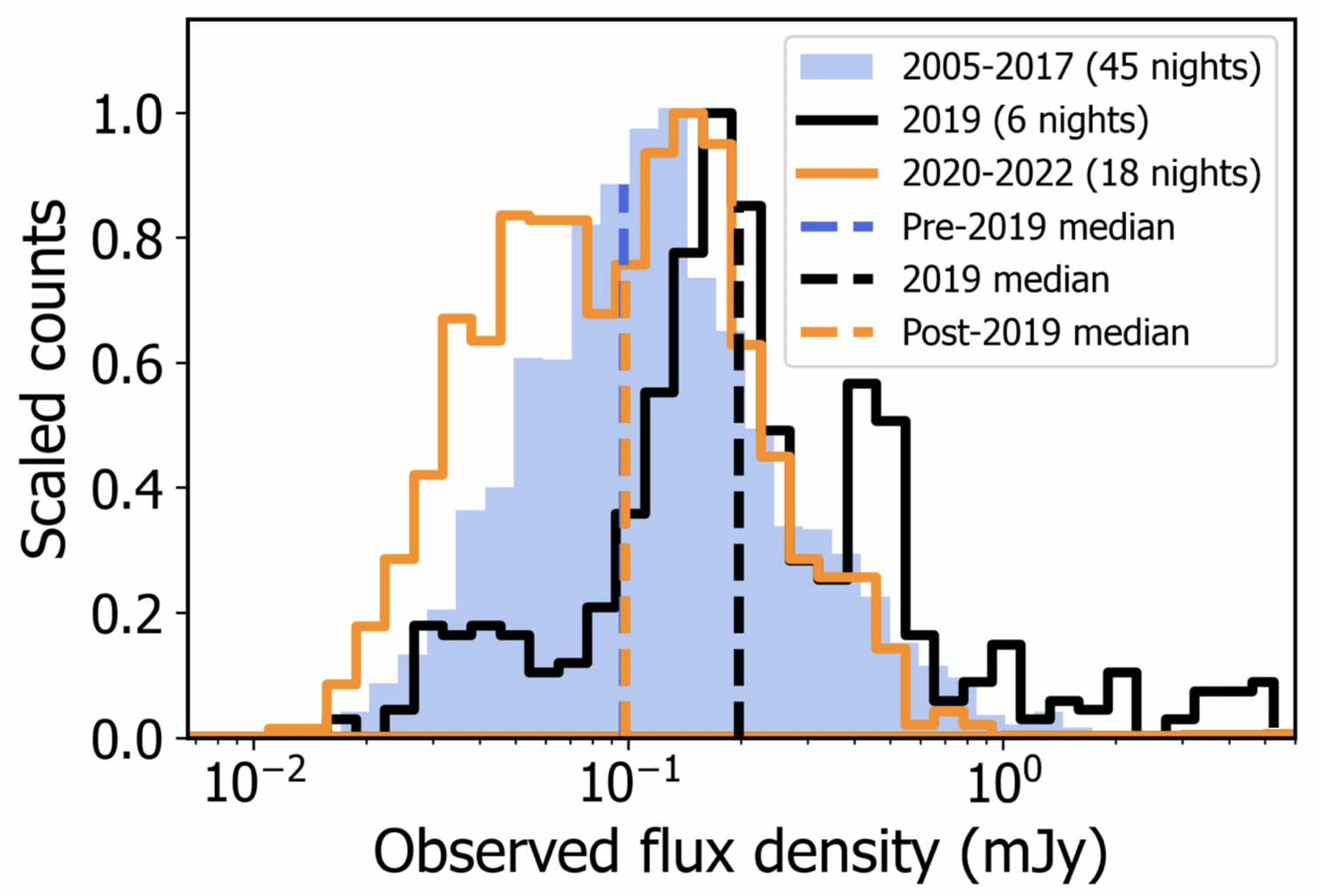}

\caption{\footnotesize Measured flux distributions of Sgr A* scaled to compare peaks and shapes of the pre-2019 (blue), 2019 (black), and post-2019 (orange) distributions, with dashed lines showing the distribution medians. We find that the 2019 flux distribution is elevated with respect to the pre- and post-2019 distributions.}
\label{fig:histograms}
\end{figure}

The flux distributions of Sgr A* reveal that flux densities in 2019 were elevated at both faint and bright levels, and that the post-2019 activity of Sgr A* is statistically consistent with the pre-2019 activity. We construct histograms of the light curves and present the flux distributions for the pre-2019 (2005-2017), 2019, and post-2019 (2020-2022) epochs in Figure \ref{fig:histograms}. In our sample, Sgr A* is detected $\sim$93\% of the time. Our treatment of non-detections is discussed in Appendix \ref{app:nondetections}.

As shown in \cite{Do19}, the 2019 distribution has a tail extending to high flux densities that is not present before 2019. Our post-2019 results show no such extended tail, indicating that we have seen no high flux densities at 2019 levels in more recent years. We observe that the shapes of the pre- and post-2019 distributions appear to be quite similar, whereas the 2019 distribution appears skewed to higher flux densities.

We are also able to measure the long-term median of the flux distribution, enabled by our consistent reduction methodology with photometric corrections for stellar confusion. We find that the median of the 2019 distribution is 0.20 mJy (1.9 mJy dereddened), which doubles the 0.10 mJy (1.0 mJy dereddened) medians of the pre- and post-2019 distributions. Other percentiles of the flux distribution are presented in Table \ref{tab:percentiles}. We observe that in addition to the median, the flux levels of 2019 are elevated at all percentiles compared to the preceding and succeeding distributions, demonstrating that the 2019 behavior was indeed unusual beyond merely the extremely bright event of 2019-05-13. On the other hand, nearly all percentiles are similar between the pre- and post-2019 distributions. The exception is at the high flux end, where discrepancies between the pre- and post-2019 distributions are explained by a recent lack of bright flux excursions. Furthermore, by adding our flux distributions and dividing by the number of observations within each time period, we can compare the average luminosity of Sgr A* at 2.12 $\mu$m. We find that Sgr A* was on average $\sim$3 times more luminous at this wavelength in 2019 than in the pre-2019 observations. Since 2019, Sgr A* has had an average luminosity of $\sim$0.9 times its pre-2019 luminosity. 

\begin{table*}
\begin{center}
\setlength{\tabcolsep}{6pt}
\caption{Flux distribution percentiles}
\label{tab:percentiles}
\begin{tabular}{c c c c }
\hline\hline
Percentile & Pre-2019 flux & 2019 flux & Post-2019  flux\\
& (mJy) & (mJy) & (mJy) \\
\hline
5\%  & 0.01 & 0.03 & 0.03\\
14\%  & 0.03 & 0.06 & 0.04\\
25\% & 0.05 & 0.13 & 0.05\\
50\% & 0.10 & 0.20 & 0.10\\
75\% & 0.17 & 0.41 & 0.18\\
86\%  & 0.26 & 0.55 & 0.23\\
95\% & 0.47 & 1.68 & 0.40\\
\hline \hline
\end{tabular}
\end{center}
\end{table*}

We can use two-tailed Kolmogorov-Smirnov (KS) tests to quantitatively compare our flux distributions. We note that KS-tests assume independent measurements and Sgr A* light curves are time-correlated on an intra-night basis; however, as in \cite{Do19}, we can compare flux distributions with multiple nights of observations that are uncorrelated with one another to roughly compare the general behavior of Sgr A* over long time periods. We use the inferred flux distributions (described in Appendix \ref{app:nondetections}) to mitigate the bias induced by non-detections. Performing a KS-test between our pre-2019 and 2019 distributions yields a KS-statistic of 0.39 ($p \ll 1$), showing it is highly unlikely that these data sets come from the same underlying distribution. \cite{Do19} also used a KS-test and found a large disagreement between the 2019 sample and historical data. Performing a KS-test between our 2019 and post-2019 distributions yields a KS-statistic of 0.36 ($p \ll 1$), showing these data sets are also highly unlikely to come from the same underlying distribution. On the other hand, performing a KS-test between our pre-2019 and post-2019 distributions yields a much lower KS-statistic of 0.04 ($p \sim 0.03$), showing these distributions are more similar. If we restrict our KS-test to flux densities above 0.05 mJy (where non-detections become negligible), we find a comparable KS-statistic of 0.05 ($p \sim 0.01$). Based on these results, we find that the pre-2019 and post-2019 flux distributions of Sgr A* agree at the $\sim$2$\sigma$ level. 

\begin{figure}
\includegraphics[width=3.3in]{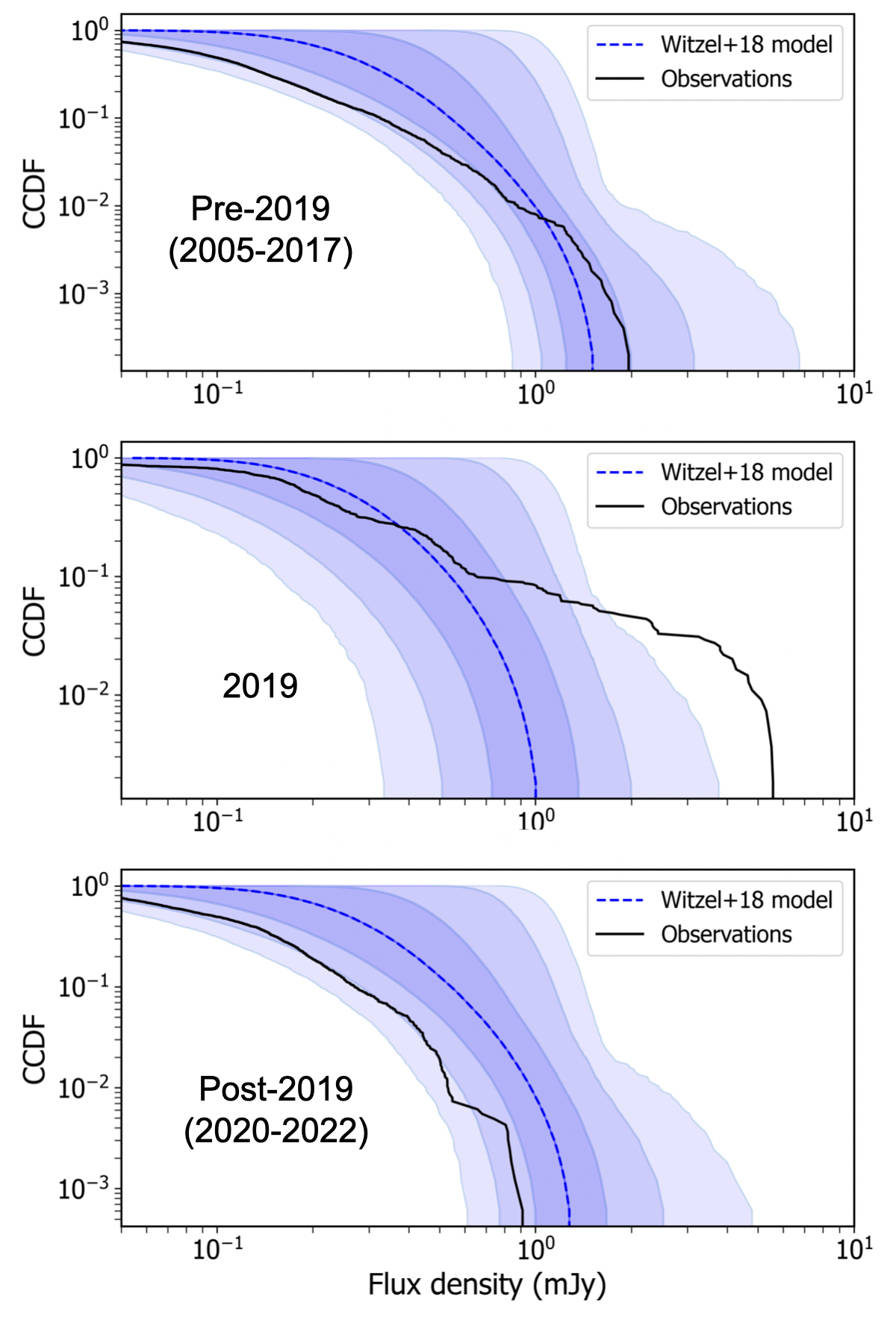}
\caption{\footnotesize Comparisons (top: pre-2019, middle: 2019, bottom: post-2019) of the complementary cumulative distribution function (CCDF) of observed data (solid black lines) with the median (blue dashed lines) and 1$\sigma$, 2$\sigma$, and 3$\sigma$ intervals (blue shaded contours) of 10,000 simulated light curves with the respective time sampling. The parameters to generate the simulated light curves were drawn from the log-normal, red-noise posterior of \cite{Witzel18}. These simulations show that while the 2019 data are inconsistent with model predictions at the $>$3$\sigma$ level, the pre- and post-2019 activity generally agrees with the model at the 1-2$\sigma$ level. We note that our 2019 data includes additional observations from late 2019 that were not incorporated into the analysis of \cite{Do19}.}
\label{fig:witzel}
\end{figure}

In addition to consistency between the pre-2019 and post-2019 flux distributions, we find that the post-2019 data agree with model predictions. As in \cite{Do19}, we compare the observations to the model 3 of \cite{Witzel18}, which is a log-normally distributed red noise process. This model was informed by over 13,000 infrared measurements from Keck, VLT, and Spitzer and importantly accounts for the temporal correlations of Sgr A*'s flux with time. For each of our three samples (pre-2019, 2019, and post-2019), we draw 10,000 parameter combinations from the posterior and for each parameter combination, generate a single light curve with the time sampling of the nights in the respective sample (45 nights for pre-2019, 6 nights for 2019, and 18 nights for post-2019). In Figure \ref{fig:witzel}, we present a comparison of the observed data with the median, 1$\sigma$, 2$\sigma$, and 3$\sigma$ credible intervals of the complementary cumulative distribution function (CCDF $\equiv$ 1 - CDF) for each set of 10,000 simulated light curves.  

Consistent with the expectation that our historical sample should agree with the established model, our pre-2019 observations match the historical model at the 1-2$\sigma$ level. We find that the 2019 bright flux excursions fall outside of the model's 3$\sigma$ intervals, as shown in \cite{Do19}, but with new observations from late 2019. Of most interest to this study is the behavior of Sgr A* in the years since 2019. We find that the post-2019 observations fall within the predictions of the historical model. The lack of high fluxes in more recent years does place these observations below model predictions at the $\sim$2.5$\sigma$ level (for fluxes reaching $\sim$0.6 mJy, i.e. all the data except the two brightest events) and at the $\sim$1.5$\sigma$ level for the highest measured flux densities. While additional observations in coming years may reveal a more statistically significant lack of activity, our current results demonstrate that the recent observations are consistent with historical expectations at the $\sim$2$\sigma$ level. Sgr A* has displayed statistically typical, albeit diminished levels of activity.

\subsubsection{Impact of the brightest night on the flux distribution}
\label{subsubsec:2019b}

We examine the impact that the night with the brightest event (2019-05-13) has on our results by constructing the 2019 flux distribution without this night. This investigation is motivated by the possibility of 2019 being an otherwise ordinary era with a single rare event that biases the flux distribution to higher levels. The flux distribution and percentiles with this night removed can be found in Appendix \ref{app:2019b}. We find that the removal of this night certainly impacts the high end of the flux distribution, but the median of the distribution remains the same and the peak of the distribution remains shifted to higher flux levels with respect to the historical distribution. Although the other nights in 2019 individually do not reach unprecedented levels, together they display a relative concentration of high flux densities. A KS-test between the pre-2019 distribution and the 2019 distribution without the brightest night yields a KS-statistic of 0.42 ($p \ll 1$), reinforcing that these data sets are highly dissimilar. As such, we conclude that our finding that the 2019 flux distribution is significantly different is robust whether or not the brightest night is included. The inclusion of this night only bolsters the result.

\subsection{Bright events in the NIR Sgr A* light curve}

\begin{figure*}[t]
\includegraphics[width=7in]{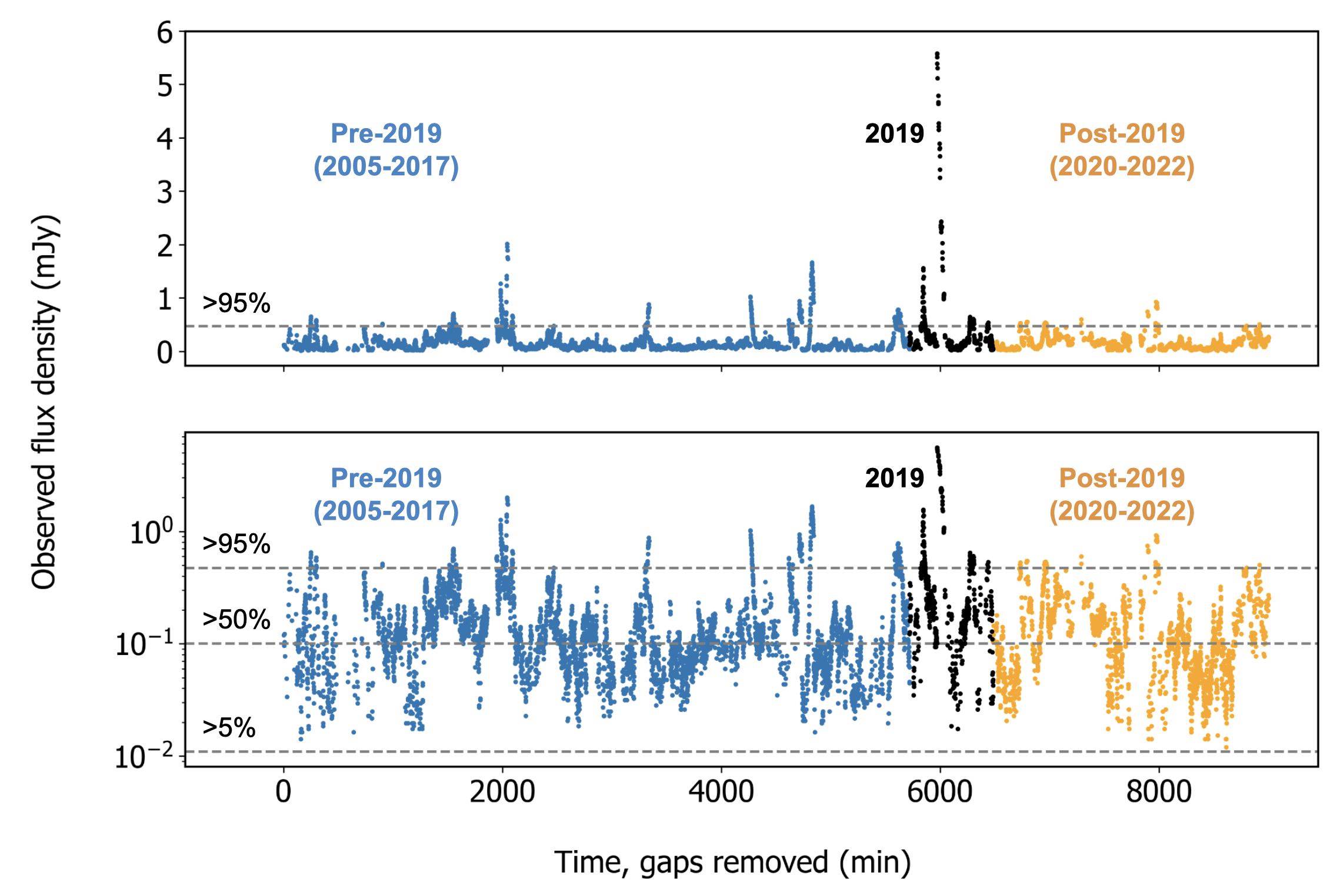}
\caption{\footnotesize Light curve of Sgr A* containing all observations in this study with time gaps between epochs removed, displayed in linear space (top) and logarithmic space (bottom). The observations are separated into pre-2019 (blue), 2019 (black), and post-2019 (orange). Dashed lines show the percentage of fluxes fainter than that level in the pre-2019 sample. Only detections that cross zero with $>$1$\sigma$ significance are shown. The bottom $\sim$5\% of frames are not displayed due to either being not detected at all or not detected with $>$1$\sigma$ significance.}
\label{fig:lightcurve}
\end{figure*}

The light curve of Sgr A* over 17 years shows the heightened flux densities in 2019 and relative lack of activity in the following years (see Figure \ref{fig:lightcurve}). To construct this light curve, we stitch together the pre-2019, 2019, and post-2019 data and remove time gaps between observation epochs. Individual light curves from each of these three subsets displaying both bright and faint flux densities are shown in Appendix \ref{app:lightcurves}.  

We find no bright fluxes in the post-2019 data at the level of the two brightest events in 2019. In six nights of 2019 observations, the two brightest events have maximum flux densities of 5.58 $\pm$ 0.04 mJy (corresponding to 53.8 mJy dereddened) and 1.56 $\pm$ 0.03 mJy (15.0 mJy dereddened). In seven nights of observations in 2020 and 2021, the highest flux density measured is only 0.60 $\pm$ 0.04 mJy (5.8 mJy dereddened). In 11 observation epochs in 2022, we observed only two bright flux excursions with maximum flux densities of 0.74 $\pm$ 0.06 mJy (7.1 mJy dereddened) and 0.93 $\pm$ 0.03 mJy (9.0 mJy dereddened). Neither of these are comparable to the 2019 events.

There is also a lack of bright flux excursions in the post-2019 data as compared to the pre-2019 data. We observe four events in the pre-2019 data (counting two on 2010-05-05) with flux densities greater than 1 mJy and no flux densities at this level post-2019. Six events with flux densities greater than 0.8 mJy are observed in the pre-2019 data, and only one is observed post-2019. The post-2019 sample is only $\sim$43\% as long as the pre-2019 sample, but there remains a relative absence of bright fluxes in more recent years.

\subsection{Sgr A* has no quiescent state in the NIR}

Our new methodology allows for measurements of faint flux densities in the NIR over 17 years and the ability to search for a quiescent, quasi-steady state. Such a state would present itself as a "floor" to the Sgr A* light curve. This study is the first to examine faint Sgr A* flux densities from 2005-2022 while making corrections for stellar confusion, allowing for reliable comparisons of faint Sgr A* flux densities over a long time baseline to search for such a floor.

We find that there is no quasi-steady floor to the NIR Sgr A* light curve at faint flux densities. When viewed in linear flux space (such as the upper light curve in Figure \ref{fig:lightcurve}), Sgr A* does appear to spend most of its time in a faint state that is punctuated by bright "flares". However, the data show that Sgr A* has been highly variable even at low flux densities over 17 years. By examining the light curve in logarithmic flux density space (bottom light curve in Figure \ref{fig:lightcurve}), we see that Sgr A* shows no quiescent state in the NIR and instead displays stochastic variability over a factor of $\sim$500 in flux. Flux densities do not level off at some faint value, but are continuously and significantly variable down to about 0.01 mJy (0.1 mJy dereddened), which is the limit where we are able to detect Sgr A* with $>$1$\sigma$ confidence and which only a small fraction of frames likely fall below.

\subsection{Timing characteristics of the light curves}

\begin{figure}
\includegraphics[width=3.3in]{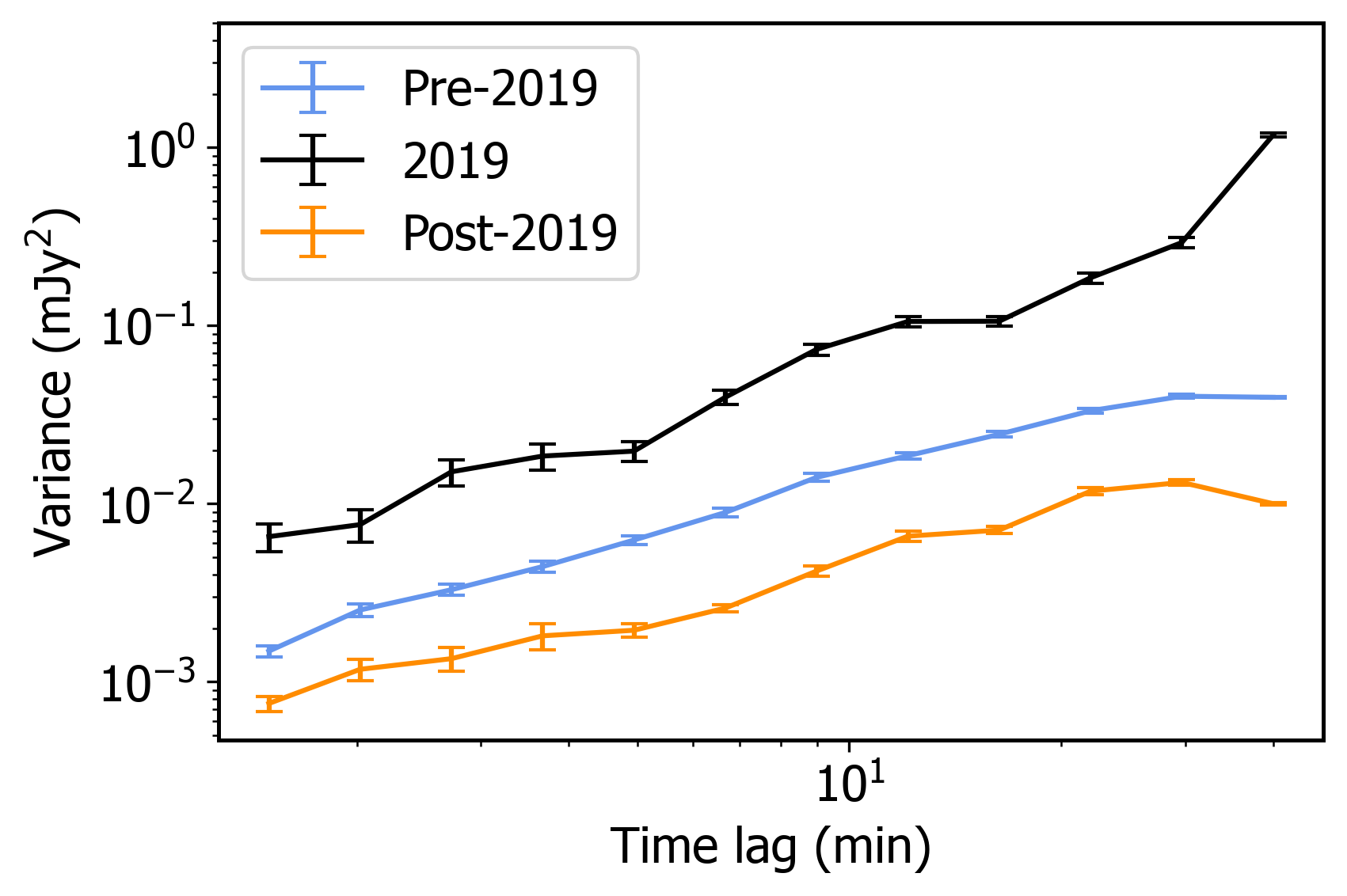}
\caption{\footnotesize Structure functions of the pre-2019 (blue), 2019 (black), and post-2019 (orange) light curves. We see that the 2019 light curves have higher variance than the pre-2019 light curves, and the post-2019 light curves have lower variance than the pre-2019 light curves.}
\label{fig:structurefunctions}
\end{figure}

We can use the first order structure function to examine the timing characteristics of the NIR light curves, as is often done in timing analyses for both Sgr A* and extragalactic AGN \citep[e.g.][]{Simonetti78, Hughes92, Do09}. The structure function ultimately allows one to determine the power spectral density (PSD) slope for a set of unevenly sampled data. For a light curve with flux measurements $F(t)$ at times $t$, the first order structure function $V(\tau)$ is defined as
\begin{equation}
    V(\tau) \equiv \langle [F(t + \tau) - F(t) ]^2 \rangle.
\end{equation}
We bin the time lags $\tau$ and distribute the $V(\tau)$ values into the corresponding bins. Logarithmic binning is used to more evenly distribute data points into bins, as there are many more samples for smaller time lags. The average of $V(\tau)$ values within each bin is used as the structure function value for that bin. The error associated with each bin is $\sigma_{\text{bin}}$/$\sqrt{N_{\text{bin}}}$, where $\sigma_{\text{bin}}$ is the standard deviation of $V(\tau)$ values within that bin and $N_{\text{bin}}$ is the number of values in the bin. The structure functions for the pre-2019, 2019, and post-2019 light curves are presented in Figure \ref{fig:structurefunctions}.

The logarithmic slope of the structure function $\beta$ (where $V(\tau) \propto \tau^{\beta}$) can be related to the power law index $\alpha$ of the PSD ($P \propto f^{-\alpha}$) for each of the pre-2019, 2019, and post-2019 samples. To make the conversion from measured measured $\beta$ to $\alpha$, we generate $10^3$ light curves with the respective time sampling and fixed PSD slope $\alpha$, then measure the structure function slope $\beta$. This procedure is repeated for varying values of $\alpha$ between $\alpha=1.5$ and $\alpha=3.0$. We perform a linear fit between $\alpha$ and $\beta$ (see Appendix \ref{app:structurefuncs}), then use this linear fit to convert the structure function slopes we measure in the real data to PSD slopes. The measured values of $\beta$ and $\alpha$ for the data are given in Table \ref{tab:PSDfits}. The structure function slope $\beta$ is measured in the regime where the structure function is approximately linear (5-40 minutes) to avoid artifacts from three-minute dithering and increased white noise in the observations at short time lags. However, we do factor a systematic uncertainty into our reported $\alpha$ and $\beta$ measurements computed from the standard deviation of structure function fits in several time intervals (1-40 minutes, 3-40 minutes, 5-40 minutes, 1-30 minutes, 3-30 minutes, and 5-30 minutes).

\begin{table*}
\begin{center}
\setlength{\tabcolsep}{6pt}
\caption{PSD fits from the structure function, where the structure function $V(\tau) \propto \tau^{\beta}$ and the PSD $P \propto f^{-\alpha}$}
\label{tab:PSDfits}
\begin{tabular}{c c c }
\hline\hline
Observation Years & $\beta$ & $\alpha$ \\
\hline
2005-2017  & 0.99 $\pm$ 0.08 & 2.35 $\pm$ 0.14\\
2019  & 1.49 $\pm$ 0.17 & 2.74 $\pm$ 0.30\\
2020-2022 & 1.03 $\pm$ 0.15 & 2.20 $\pm$ 0.19\\
\hline \hline
\end{tabular}
\end{center}
\end{table*}

We see that the structure functions have PSD slopes of $\alpha \approx$ 2-3, consistent with previous findings \citep[e.g.][]{Do09} that Sgr A* displays red-noise, power-law behavior with higher variability at longer time lags.  We find a steeper $\alpha$ for 2019 and a shallower $\alpha$ for the post-2019 data. However, it is worth noting that a more significant white noise component from stellar contamination in 2020-2022 could cause us to measure a shallower slope for these years (see Appendix \ref{app:structurefuncs}). Interestingly, the 2019 structure function is significantly elevated with respect to the others at all time lags, indicating a higher variance at every time lag. On the other hand, the post-2019 structure function falls below the others at all time lags, indicating lower variance at every time lag. Our structure functions reveal that greater average luminosity in 2019 correlates with more extreme variability, and the lower average luminosity post-2019 correlates with relatively diminished variability.

\section{Discussion}
\label{sec:physics}

Our results show that the 2019 flux distribution was shifted to higher levels and had larger variance. In 2020-2022, the behavior of Sgr A* was statistically similar, but slightly diminished compared to that of 2005-2017. We also find that Sgr A* has no quiescent state over 17 years of NIR observations. These findings have implications for understanding the time-variable accretion flow onto Sgr A*.

The pronounced shift in the 2019 median flux density suggests that Sgr A* experienced a heightened accretion rate, revealing that the 2019 NIR activity was extraordinary beyond just the bright events. \cite{GRAVITY20} also reported an elevated median in 2019 (compared to the years 2017 and 2018), although it was found not statistically significant given the number of observations over a time baseline of 3 years. Our time baseline of 17 years provides stronger evidence for a shift in the median of the flux distribution in 2019. The two-state model for the flux distribution with a log-normal quiescent state and power-law tail \citep{Dodds-Eden11, GRAVITY20} interprets bright flares as manifestations of such a tail, but does not account for the changes to the faint emission that we observe in 2019 in this study. Because we see elevated activity at both low and high levels in 2019, explanations invoking a physical change to Sgr A*'s accretion state are favored over a new statistical model.

Multi-wavelength observations of Sgr A* in 2019 also favor a physical change to the accretion flow. The temporal concentration of high fluxes observed in 2019 by Keck and VLT in the NIR, as well as heightened activity in radio \citep[e.g.][]{Murchikova21b, Boyce22, Cheng23} and X-rays \citep[e.g.][]{Pavlinsky19, Degennar19} is also evidence for a physical disturbance during 2019. If Sgr A*'s variability were better characterized by a different underlying distribution than that of \cite{Witzel18} (a single stochastic process with a log-normal distribution), we would expect to see bright events and elevated multi-wavelength activity more evenly distributed over two decades, rather than clustered together in a single year. 

Our results favor models in which extra gas was deposited onto the black hole in 2019, temporarily increasing the accretion rate and stimulating the production of frequent bright events. Such models include a disturbance induced by the closest approach of the star S0-2 in 2018 or the delayed infall of gas pulled from a tidal interaction with the dusty object G2 in 2014. \cite{Kawashima17} predicted a time-delayed radio and infrared brightening of Sgr A* around 2020 caused by the passage of G2. In their simulations, the magnetic energy within the accretion disk increases by a factor of 3-4 after $\sim$5 years following the passage of G2. \cite{Murchikova21} calculates that the infall time of material from G2 (and the possible simultaneous infall of material from the object G1) more closely matches the time delay between pericenter passage and flaring time than that of S0-2. Hydrodynamic simulations have also shown that winds from the passage of S0-2 are unlikely to have a measurable effect on the inner accretion flow \citep{Ressler18}. If the 2019 activity was indeed caused by a temporary accretion increase, these findings collectively favor the G2 hypothesis over the S0-2 hypothesis as a source for the excess material.

We rule out a long-term elevated accretion state, as the variability of Sgr A* resembles its past activity and no extraordinarily bright events have been observed in more recent years. In fact, we have noted a slight lowering of the post-2019 flux distribution at the high end, although this effect is not statistically significant. Numerical general relativistic magnetohydrodynamical (GRMHD) simulations have revealed that bright flares powered by magnetic reconnection can eject part of the accretion disk and suppress the mass accretion rate onto the black hole \citep{Ripperda22}. Changes to the magnetic flux content of the disk in simulations have measurable effects on the light curves and flux distributions of Sgr A* \citep{Chatterjee21}. The lack of high flux densities post-2019 and the slight decrease in mean luminosity could imply that the 2019 activity altered the accretion flow onto Sgr A* for an extended period of time. With more observations, we will be able to determine whether the relative inactivity of Sgr A* in more recent years is in fact statistically significant.

A major challenge to precision measurement of the flux distribution of Sgr A* is stellar confusion. Recent observations with NIR interferometry \citep[e.g.][]{GRAVITY20} have improved angular resolution and reduced the impact of confusion on flux measurements, but such observations of Sgr A* have only begun in 2017. Single-telescope data is complementary to interferometric data because it offers a long time baseline to compare the behavior of Sgr A* across decades, but as in this work, care must be taken to account for confusion. Future work to combine single-telescope and interferometric data sets would help reduce systematics in the study of Sgr A* variability.

\section{Conclusion}
\label{sec:conclusion}

Each hypothesis for the 2019 activity of Sgr A* is scientifically plausible for understanding the activity in terms of an excess accretion flow onto the black hole. If we are able to trace the activity to S0-2 or G2, we will directly connect the supermassive black hole's feeding behavior to an object in its vicinity and learn more about how neighboring objects supply material for the black hole to accrete. If Sgr A* has undergone a change of state, we will catch a real-time glimpse into the physical response of a supermassive black hole accretion flow to an increase in gas flow. In any case, it is important to determine whether the 2019 activity is representative of Sgr A*'s typical behavior or a period of extraordinary activity. If the former, the 2019 activity should be included in long-term models of Sgr A*. If the latter, the unusual behavior should be treated as a separate period with distinct physical characteristics. 

To investigate the origin of the 2019 activity, we have done the following:
\begin{itemize}
    \item Performed a consistent reduction of Keck Sgr A* observations from 2005-2022 with new treatment for stellar confusion to enable a robust comparison of flux densities over nearly two decades
    
    \item Demonstrated that both faint and bright 2019 flux densities were significantly elevated with respect to pre- and post-2019 observations

    \item Shown that the average luminosity of Sgr A* was $\sim$3 times higher than historical measurements in 2019 and $\sim$0.9 times historical measurements in the years 2020-2022

    \item Shown that while the post-2019 observations are statistically consistent with the historical sample, as well as with the statistical model of \cite{Witzel18}, we have observed a relative deficiency in activity in 2020-2022 at the $\sim$2$\sigma$ level

    \item Demonstrated that 2019 showed heightened variability with respect to the years before and after, indicating that a greater average luminosity seems correlated with more extreme variability

    \item Found no evidence for a NIR quiescent state of Sgr A*; Sgr A* displays stochastic flux variations over a factor of $\sim$500 and is continuously variable down to about 0.01 mJy
\end{itemize}

 We argue that the observed increase in NIR flux densities in 2019, along with the concentration of bright events measured by many observatories that year, points to a transient increase in accretion activity during 2019. In these models, excess gas is temporarily deposited onto the black hole from the closest approach of a nearby object, raising the median flux and increasing the probability of bright events. Numerical simulations and analytic calculations have favored the black hole's tidal interaction with G2 in 2014 as a source for this excess material. Further modeling of G-object interactions with the accretion flow and their ability to cause the 2019 activity should be done to provide a more robust comparison to observations.

We have ruled out the hypothesis that 2019 is a precursor to a long-term elevated accretion state, as the current behavior of Sgr A* resembles its behavior before 2019, and we have seen a lack of high fluxes in more recent years. Future NIR observations will continue to monitor the black hole for signs of increased (or decreased) activity.  Further analyses of multi-wavelength data will help us compare the physical mechanisms behind the bright events of 2019 and other years. 

We thank the anonymous referee for their helpful comments. We thank the staff and astronomers at the Keck Observatory for their help in taking the observations, especially Jim Lyke, Randy Campbell, Percy Gomez, Carlos Alvarez, Greg Doppmann, Michael Lundquist, Rosalie McGurk, Joel Aycock, Tony Connors, John Pelletier, Julie Renaud-Kim, Arina Rostopchina, Heather Hershley, and Tony Ridenour. The W. M. Keck Observatory is operated as a scientific partnership among the California Institute of Technology, the University of California, and the National Aeronautics and Space Administration. The Observatory was made possible by the generous financial support of the W. M. Keck Foundation. The authors wish to recognize that the summit of Maunakea has always held a very significant cultural role for the indigenous Hawaiian community. We are most fortunate to have the opportunity to observe from this mountain.

\software{
    NumPy \citep{numpy},
    SciPy \citep{scipy},
    Matplotlib \citep{matplotlib},
    KAI \citep{KAI}
}

\bibliography{paperbib, softwarebib}{}
\bibliographystyle{aasjournal}

\appendix

\section{Photometry}
\label{app:reduction}

\begin{figure*}[ht]
\centering
\includegraphics[width=4in]{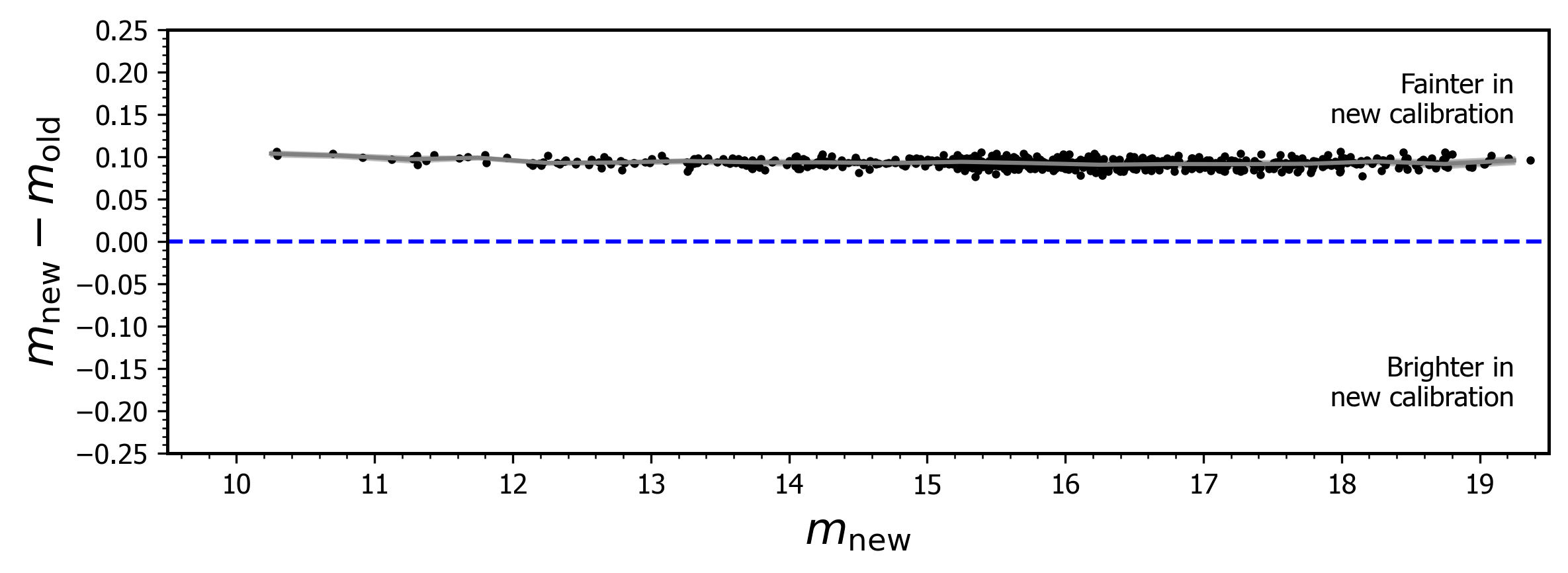}
\caption{\footnotesize Comparison of observed magnitudes for Galactic center sources between the photometric calibration of \cite{Do19} and the new calibration in this work (Gautam et al. in prep.), as a function of magnitude in the new calibration. We see that sources are systematically fainter by about 10\% with the new calibration.}
\label{fig:calibrations}
\end{figure*}

We updated the reference fluxes for calibrator stars from those that were used in \cite{Do19} and \citet{Gautam19}. The magnitude difference between the previous and new photometric calibrations for sources in the central 10 arcseconds is $0.093 \pm 0.003$ and in the central 1 arsecond in $0.096 \pm 0.003$, with stars being systematically fainter in the new calibration (see Figure \ref{fig:calibrations}). In this work, we used the following calibrator stars: IRS~16NW, S3-22, S1-17, S1-34, S4-3, S1-1, S1-21, S3-370, S3-88, S3-36, and S2-63. We derived reference flux measurements for the selected calibrator stars using the \citet{Schodel10} photometric catalog transformed to the Keck NIRC2 bandpasses. The procedure to transform fluxes from VLT NACO $K_{s}$ and $H$-bands to Keck NIRC2 $K'$ and $H$-bands is described in more detail in \citet{Gautam19}. More details of our new photometric calibration are forthcoming in Gautam et al. in prep.

\begin{figure*}
\centering
\includegraphics[width=4in]{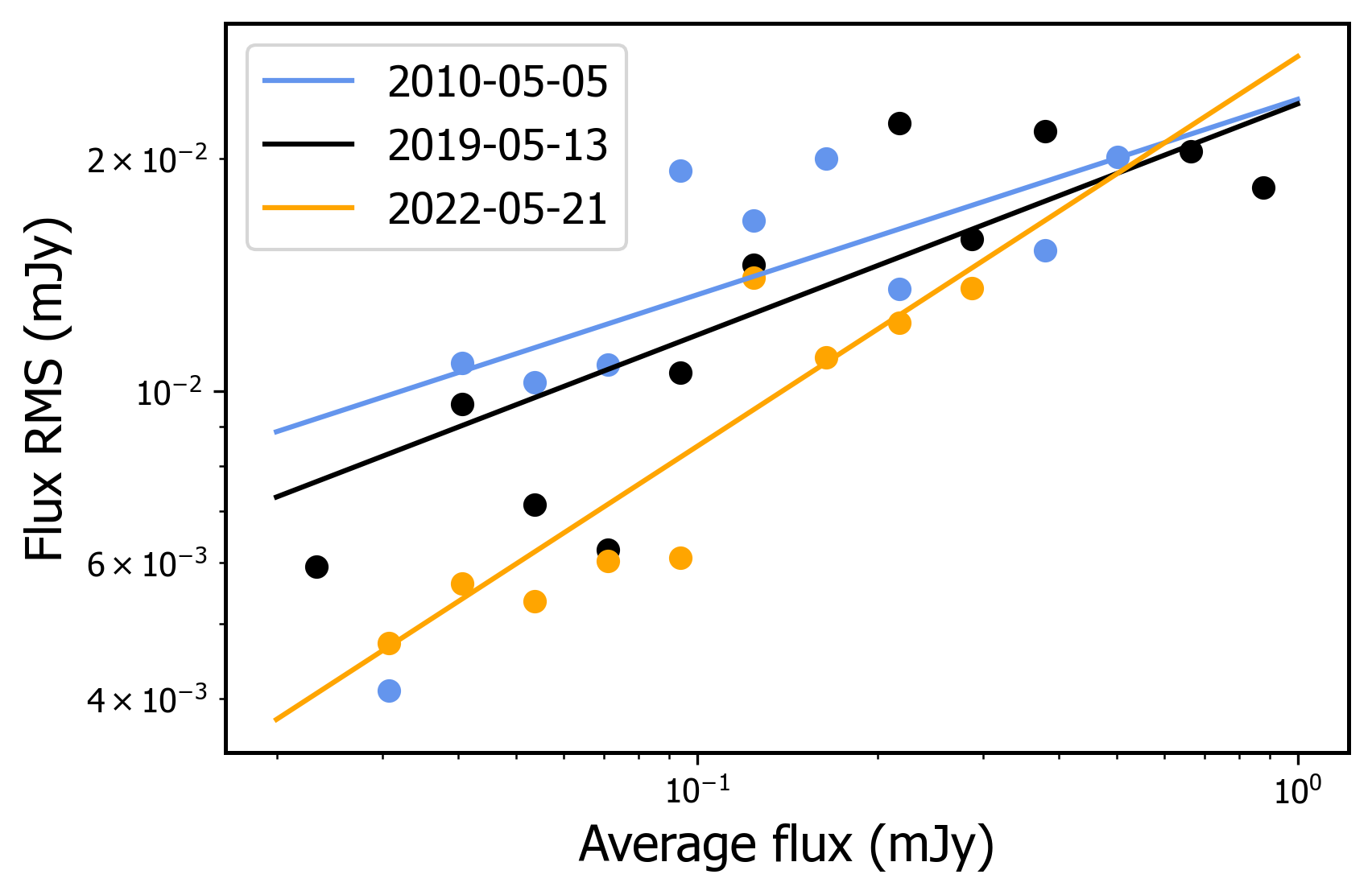}
\caption{\footnotesize Sample power law fits to the RMS fluxes for stars within 1 arcsecond of Sgr A* within the typical Sgr A* flux range. Shown here are the fits for three observing epochs (2010-05-05, 2019-05-13, 2022-05-21). We perform this procedure for each of the 69 nights in our sample and use the respective power law fits to determine the uncertainties on the Sgr A* flux measurements.}
\label{fig:uncertaintyfit}
\end{figure*}

Photometric uncertainties were estimated by fitting a power law between the flux level and RMS flux uncertainty for stars within 1 arcsecond of Sgr A*, as described in \cite{Do09}. We then used this relationship to calculate the uncertainty in flux measurements for Sgr A*. The photometric uncertainties are generally less than 15\% for low flux values ($\lesssim$0.2 mJy) and less than 5\% for high flux values ($\gtrsim$0.2 mJy). Sample power law fits to determine flux measurement uncertainties are shown in Figure \ref{fig:uncertaintyfit}. The flux uncertainties for central arcsecond sources are not significantly different between the previous and new photometric calibrations.

\section{Photometric Stability}
\label{app:stability}

\begin{figure*}
\centering
\includegraphics[width=6in]{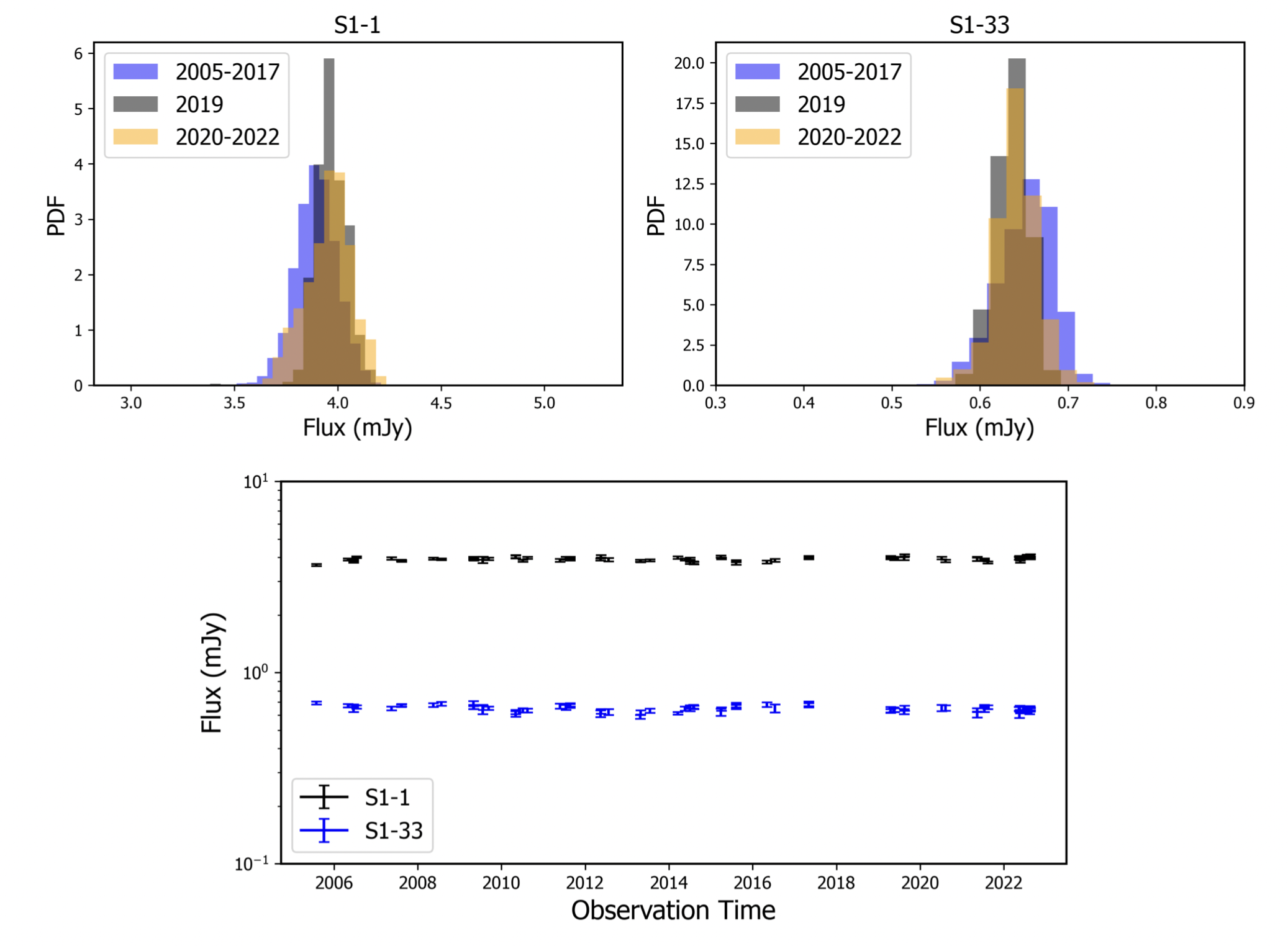}
\caption{\footnotesize Top: Pre-2019 (blue), 2019 (black), and post-2019 (orange) flux distributions of the non-variable stars S1-1 (left) and S1-33 (right). Bottom: Nightly mean and standard deviation of S1-1 (black) and S1-33 (blue) from 2005-2022. The inter- and intra-night scatters of S1-1 and S1-33 are less than $\sim$5\%, indicating we are able to consistently reproduce flux densities at this level.}
\label{fig:star_fluxes}
\end{figure*}

We investigate the photometric stability of our observations by constructing flux distributions for two comparison stars, S1-1 and S1-33, that are known to not be variable and span nearly an order of magnitude in flux (Gautam et al. in prep). The pre-2019, 2019, and post-2019 flux distributions of these stars and the nightly mean and standard deviation of the fluxes are presented in Figure \ref{fig:star_fluxes}. We find that the differences between the means and standard deviations of the flux distributions of these stars is not significant between the pre-2019, 2019, and post-2019 eras. The total 2005-2022 flux distribution for S1-1 can be described by a Gaussian with a width of $\sim$3\%, and that of S1-33 can be described by a Gaussian with a width of $\sim$5\%. The scatter of mean flux levels between individual nights for S1-1 is $\sim$2\% and for S1-33 is $\sim$4\%. Within individual nights, the average scatter is $\sim$1\% for S1-1 and $\sim$3\% for S1-33. As such, between all of our observations we are able to consistently reproduce the fluxes of S1-1 and S1-33 at the level of a few percent. The fluctuations in flux density for these non-variable stars are much smaller than the variations in Sgr A* flux density that we find, indicating that the shifts in the Sgr A* flux distributions that we report are photometrically robust.

\section{Observation Epochs Omitted}
\label{app:omissions}

In Section \ref{subsec:sample}, we describe the successive cuts made to our sample to ensure robust photometry and comparisons of Sgr A* activity over time. Out of the 104 Galactic center $K'$ observing epochs taken with Keck from 2005-2022, the 35 that we have omitted from our analysis are given in Table \ref{tab:omissions}, along with the reasons for omission.

\begin{table*}
\begin{center}
\setlength{\tabcolsep}{8pt}
\caption{Galactic center $K'$ epochs omitted from analysis}
\label{tab:omissions}
\begin{tabular}{c c}
\hline\hline
Date (UT) & Reason for Omission\\
\hline
2005-07-30 & $<20$ quality frames\\
2006-05-02 & $<20$ quality frames\\
2006-05-21 & $<20$ quality frames\\
2014-03-19 & $<20$ quality frames\\
2014-04-18 & $<20$ quality frames\\
2014-04-19 & $<20$ quality frames\\
2014-05-12 & $<20$ quality frames\\
2014-08-03 & $<20$ quality frames\\
2014-08-05 & $<20$ quality frames\\
2015-04-01 & $<20$ quality frames\\
2015-05-14 & $<20$ quality frames\\
2016-07-12 & $<20$ quality frames\\
2017-07-18 & S0-2 in confusion limit\\
2017-07-27 & S0-2 in confusion limit\\
2017-08-08 & S0-2 in confusion limit\\
2017-08-09 & S0-2 in confusion limit\\
2017-08-10 & S0-2 in confusion limit\\
2017-08-11 & S0-2 in confusion limit\\
2017-08-23 & S0-2 in confusion limit\\
2017-08-24 & S0-2 in confusion limit\\
2017-08-26 & S0-2 in confusion limit\\
2018-03-17 & S0-2 in confusion limit\\
2018-03-22 & S0-2 in confusion limit\\
2018-03-30 & S0-2 in confusion limit\\
2018-05-19 & S0-2 in confusion limit\\
2018-05-24 & S0-2 in confusion limit\\
2019-06-25 & $<20$ quality frames\\
2019-06-30 & $<20$ quality frames\\
2019-08-18 & $<20$ quality frames\\
2021-04-29 & Saturated PSF reference stars\\
2021-08-12 & $<20$ quality frames\\
2021-08-14 & $<20$ quality frames\\
2021-08-15 & Saturated PSF reference stars\\
2021-08-21 & Saturated PSF reference stars\\
2022-07-16 & $<20$ quality frames\\

\hline \hline
\end{tabular}
\end{center}
\end{table*}

\section{Light Curve Data}
\label{app:data}

We present the 2005-2022 light curve data of Sgr A* in Table \ref{tab:data}, which is used to construct the light curve shown in Figure \ref{fig:lightcurve} and the flux distributions. The Table \ref{tab:data}  data are presented as observed in the $K'$-band without corrections for extinction and before applying the photometric corrections for stellar confusion given in Table \ref{tab:corrections}. 

\begin{table*}[ht]
\begin{center}
\setlength{\tabcolsep}{7pt}
\caption{Sgr A* light curve data}
\label{tab:data}
\begin{tabular}{c c c c c c}
\hline\hline
Observation Date & Time & Flux & Flux Error & Mag & Mag Error \\
(YYYY-MM-DD) & (UT) & (mJy) & (mJy) & ($K') $& ($K'$) \\
\hline
2005-07-31	& 06:59:49.52	& 0.180	& 0.017	& 16.451 & 0.105\\
2005-07-31	& 07:03:19.17	& 0.158	& 0.016	& 16.592 & 0.112\\
2005-07-31	& 07:06:54.02	& 0.182	& 0.017	& 16.441 & 0.104\\
2005-07-31	& 07:10:15.68	& 0.165	& 0.017	& 16.550 & 0.110\\
2005-07-31	& 07:13:43.73	& 0.141	& 0.015	& 16.717 & 0.119\\
...         \\
2022-08-20	& 07:25:55.18	& 0.281	& 0.020	& 15.970 & 0.078\\
2022-08-20	& 07:31:20.56	& 0.250	& 0.019	& 16.094 & 0.082\\
2022-08-20	& 07:32:20.23	& 0.273	& 0.020	& 15.999 & 0.079\\
2022-08-20	& 07:37:47.65	& 0.319	& 0.022	& 15.831 & 0.074\\
2022-08-20	& 07:38:47.32	& 0.304	& 0.021	& 15.882 & 0.075\\
\hline \hline
\end{tabular}
\end{center}
Note: The data are presented as observed in the $K'$-band without corrections for extinction and before applying photometric corrections for stellar confusion. Table \ref{tab:data} is published electronically in its entirety in a machine-readable format in the online journal. A portion is shown here for guidance regarding its form and content. Missing flux and magnitude values in the full table correspond to non-detections of Sgr A*.
\end{table*}

\section{Non-detections of Sgr A*}
\label{app:nondetections}

\begin{figure*}[ht]
\includegraphics[width=7in]{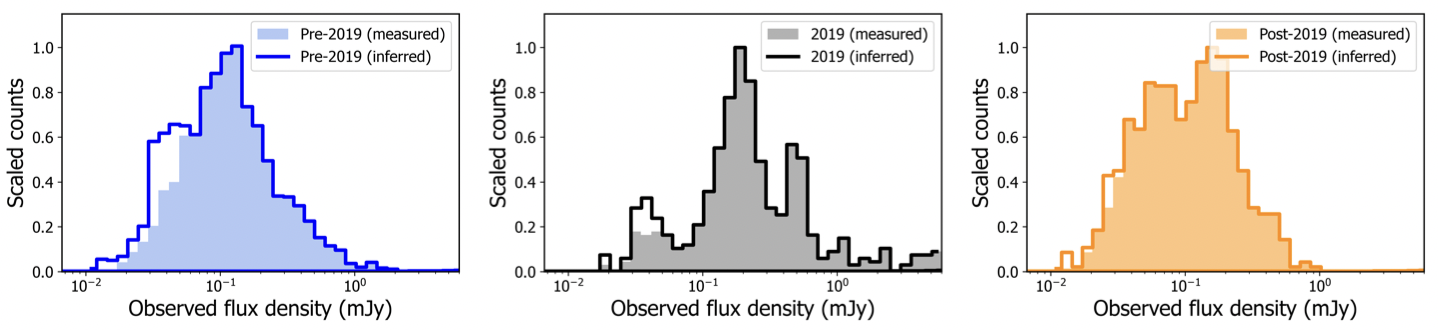}
\caption{\footnotesize } Measured and inferred flux distributions for the pre-2019 (left), 2019 (middle), and post-2019 (right) epochs. The values of non-detections in the inferred distributions are estimated from neighboring flux measurements in the light curves. 
\label{fig:inferred_fluxes}
\end{figure*}

In the data, Sgr A* is detected $\sim$93\% of the time, indicating that the flux distribution is mostly unbiased. However, because the flux of Sgr A* does not vary significantly between consecutive frames, we can estimate the values of non-detected fluxes based on the lowest neighboring flux measurement in the light curves. Comparisons between the measured and inferred flux distributions are shown in Figure \ref{fig:inferred_fluxes}. We find that Sgr A* is detected $>$99\% of the time above $F_{K_s} = 0.05$ mJy. $\sim$26\% of frames in the pre-2019 distribution, $\sim$12\% of frames in the 2019 distribution, and $\sim$24\% of frames in the post-2019 distribution fall below this threshold. As such, while we estimate flux distribution percentiles using the inferred flux distributions, the median (50th percentile) values we present are not biased by the non-detections. We do not impute values for non-detections in our timing analyses, as this procedure would artificially introduce noise.

\section{2019 Flux Distribution with Brightest Night Removed}
\label{app:2019b}

\begin{figure*}[ht]
\centering
\includegraphics[width=5in]{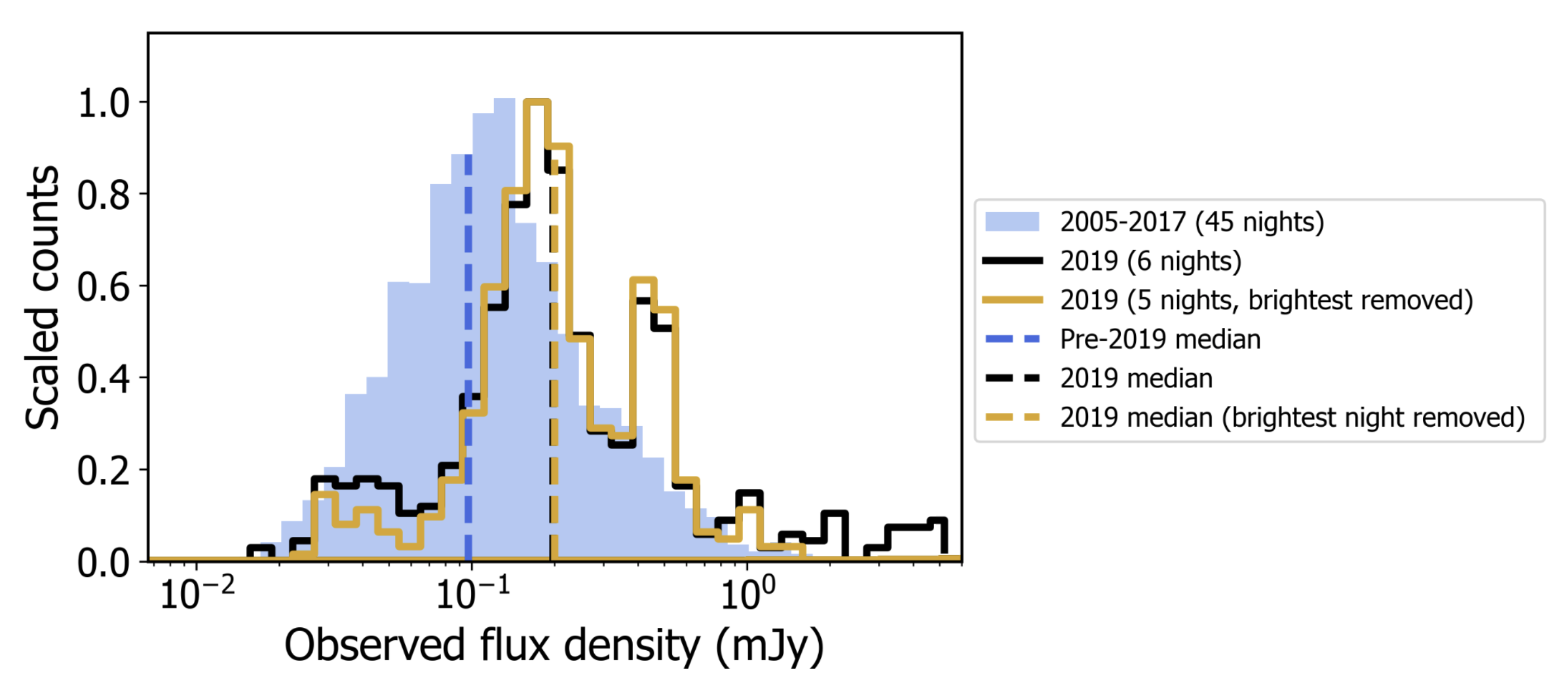}
\caption{\footnotesize  Flux distributions of Sgr A* scaled to compare peaks and shapes of the pre-2019 distribution (blue), 2019 distribution (black), and 2019 distribution with the brightest night removed (gold), with dashed lines showing the distribution medians. Although removing the brightest night lowers the high flux end of the 2019 distribution, the peak of the 2019 distribution remains elevated with respect to the historical distribution, and the 2019 median is unchanged.}
\label{fig:fluxdistribution_2019b}
\end{figure*}

In Section \ref{subsubsec:2019b}, we discuss how removing the night with the highest flux densities (2019-05-13) impacts our analysis of the flux distribution. Figure \ref{fig:fluxdistribution_2019b} shows the 2019 flux distribution without the brightest night, and Table \ref{tab:percentiles_2019b} shows the percentiles of the flux distribution.

\begin{table*}
\begin{center}
\setlength{\tabcolsep}{6pt}
\caption{Pre-2019, 2019, and 2019 (brightest night removed) flux distribution percentiles}
\label{tab:percentiles_2019b}
\begin{tabular}{c c c c }
\hline\hline
Percentile & Pre-2019 flux & 2019 flux & 2019 (brightest night removed) flux\\
& (mJy) & (mJy) & (mJy) \\
\hline
5\%  & 0.01 & 0.03 & 0.03\\
14\%  & 0.03 & 0.06 & 0.10\\
25\% & 0.05 & 0.13 & 0.13\\
50\% & 0.10 & 0.20 & 0.20\\
75\% & 0.17 & 0.41 & 0.36\\
86\%  & 0.26 & 0.55 & 0.49\\
95\% & 0.47 & 1.68 & 0.62\\
\hline \hline
\end{tabular}
\end{center}
\end{table*}

\section{Individual Light curves}
\label{app:lightcurves}

\begin{figure*}[ht]
\includegraphics[width=7in]{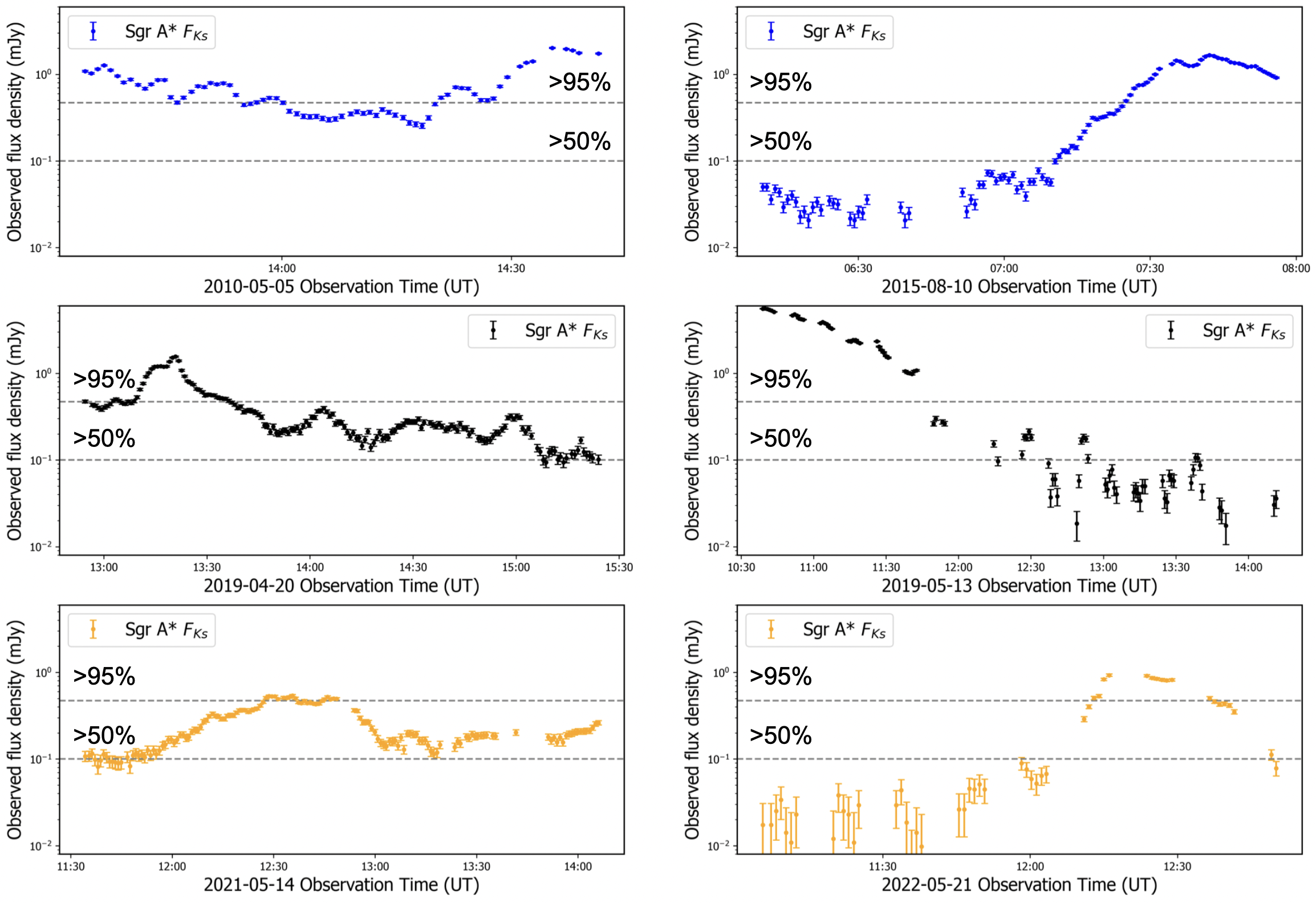}
\caption{\footnotesize } Sample observed $K_s$ light curves of Sgr A* displaying bright flux excursions. Dashed lines show the 95th and 50th percentile flux values from the historical distribution. Top row: Pre-2019 light curves with high observed flux densities. The brightest event from 45 nights of pre-2019 observations is shown on the left (2010-05-05), with a maximum flux density of $\sim$2 mJy. Middle row: Two out of six nights in 2019 far exceed historical levels. Shown on the right is the brightest ever event (2019-05-13) first reported in \cite{Do19}, with a peak flux density of $\sim$5.6 mJy more than doubling the highest pre-2019 flux density. Bottom row: Post-2019 light curves containing bright flux excursions do not reach the levels of either pre-2019 or 2019. The brightest event from 18 nights of post-2019 observations is shown on the right (2022-05-21), reaching $\sim$0.9 mJy.
\label{fig:appendix_bright_lcs}
\end{figure*}

\begin{figure*}[ht]
\includegraphics[width=7in]{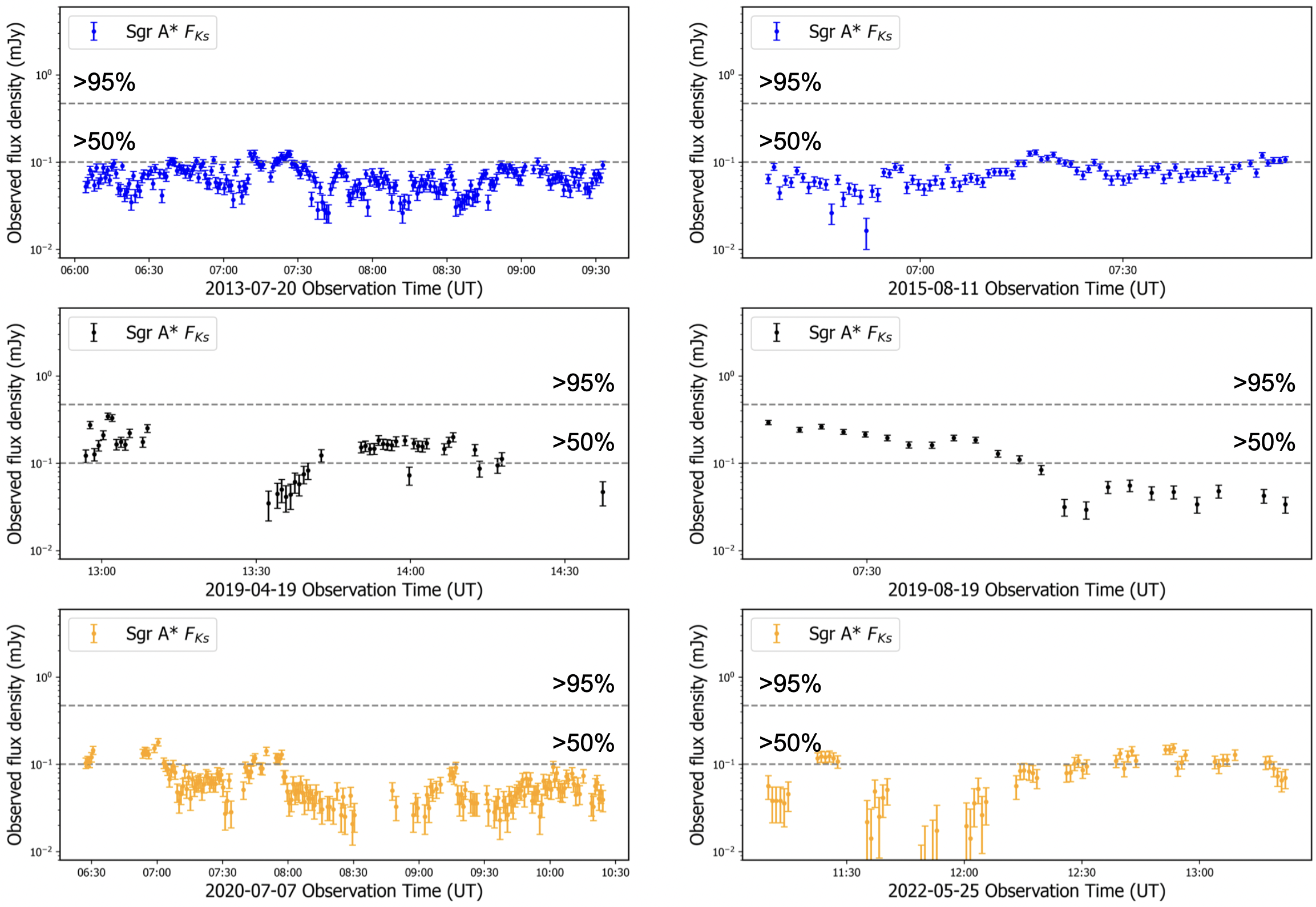}
\caption{\footnotesize } Sample of faint $K_s$ light curves of Sgr A*. Dashed lines show the 95th and 50th percentile flux values from the historical distribution. Comparing to Figure \ref{fig:appendix_bright_lcs}, we see that Sgr A* displays a range of variability at both low and high levels. Top row: Faint pre-2019 light curves. Middle row: Faint 2019 light curves. We see that even the faintest nights in 2019 display flux densities that are higher than the faintest pre- and post-2019 nights. Bottom row: Faint post-2019 light curves.
\label{fig:appendix_faint_lcs}
\end{figure*}

We present some individual light curves from pre-2019, 2019, and post-2019 observing epochs to demonstrate the range of variability that Sgr A* exhibits. Figure \ref{fig:appendix_bright_lcs} shows some of the brightest measured flux densities and Figure \ref{fig:appendix_faint_lcs} contains light curves displaying relatively low activity. 

\section{Stellar Confusion Correction}
\label{app:confusion}

The Galactic center is a crowded field with stars occasionally confusing Sgr A* on the sky, so we must consider how extended PSFs from nearby stellar sources contaminate the measured Sgr A* flux. Previous studies \citep[e.g.][]{Witzel18} subtracted yearly minima from the observed flux densities to mitigate this effect. This procedure makes all the data comparable, but deletes information about potential shifts in the median flux density caused by changes in the accretion rate. In this study, we seek to measure the shift in the median of the flux distribution between 2019 and other years to examine the accretion flow, so we take a new approach to characterizing confusion. 

\begin{figure*}[t]
\includegraphics[width=7in]{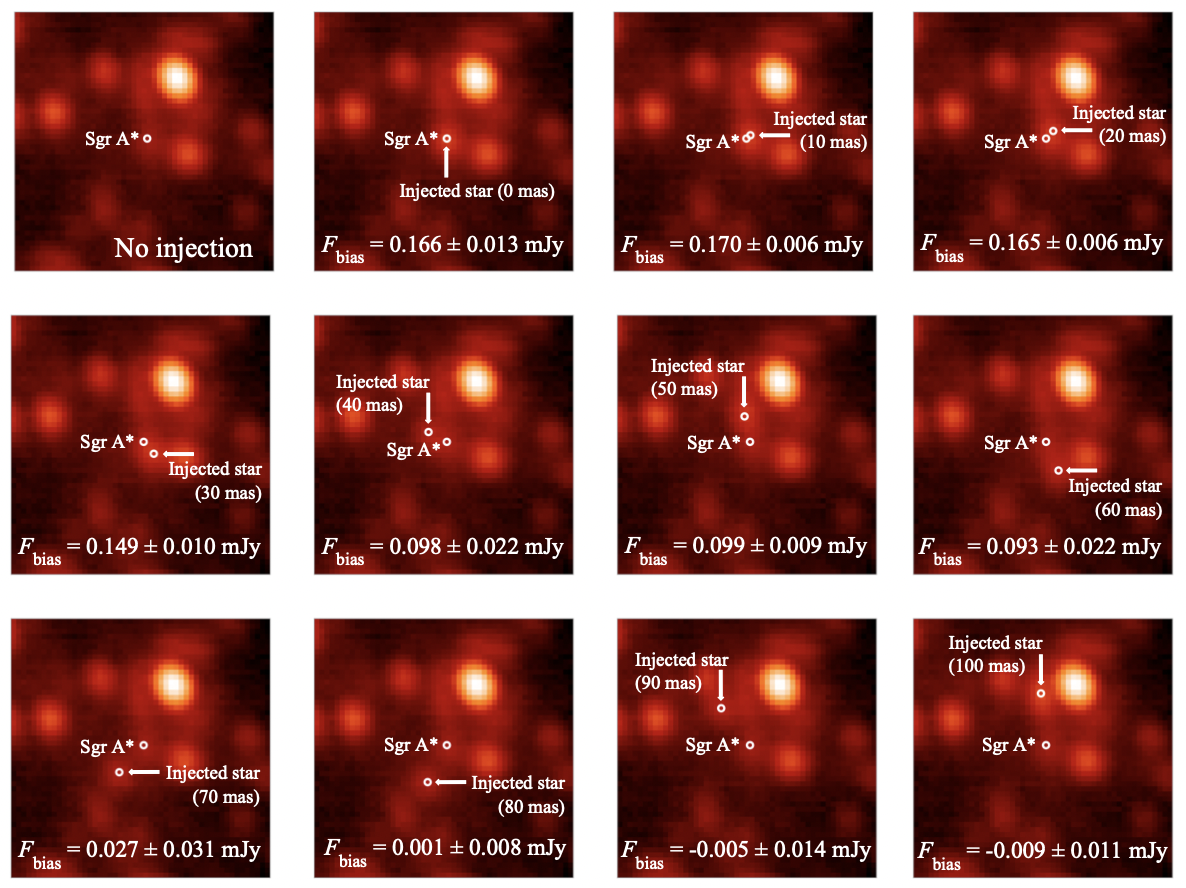}
\caption{\footnotesize } Sample images from our star injection routine. The upper left image shows the central 0.5'' $\times$ 0.5'' of a Galactic center frame from 2014-05-19 with no artificial star injected. This frame was chosen for its typical data quality (Strehl ratio $\sim$0.27), typical Sgr A* brightness of $K = 16.8$ mag, and Sgr A* being relatively unconfused by known stars. In the remaining images above, a $K = 16.5$ mag star is injected into the images with respective distances from Sgr A* labeled and random orientations. We then ran our PSF fitting program \textit{Starfinder} on each image to determine the photometric bias in our measurement of Sgr A*'s flux. This procedure was repeated for five similar frames and the average values of the photometric bias induced by the injected stars are displayed at the bottom of each image. As a general trend, the photometric bias decreases with increasing distance. We repeated this procedure for injected star brightnesses ranging from $K = 14-18$ mag. 
\label{fig:starplant_images}
\end{figure*}

To correct for the flux bias attributed to Sgr A* for epochs when a known star is within the confusion limit, we subtract a value informed from simulations of stellar confusion with Sgr A*. We injected synthetic stars near Sgr A* in a series of five unconfused images from 2014-05-19 with characteristic data quality (Strehl ratio $\sim$0.27) and median Sgr A* flux levels ($\sim$0.1 mJy), noting that our results are relatively insensitive to these choices within the uncertainties. In each set of images, we planted stars at distances from 0-100 mas (randomly oriented) and $K'$ brightnesses from 14-18 mag, then ran \textit{Starfinder} on the simulated images. See Figure \ref{fig:starplant_images} for sample images from our injection procedure. By subtracting out the known unconfused flux of Sgr A* from our detected value, we are able to obtain an estimate for the photometric bias induced by the stellar injections. The results of this procedure can be found in Table \ref{tab:starplanting}. We also measure the position of Sgr A* and report the astrometric biases in Table \ref{tab:starplanting_astrometry}. For injections below 60 mas, the source and Sgr A* are almost always detected as a combined source, making the confusion correction reliable. However, from $\sim$60-80 mas, Sgr A* and the injected star are sometimes detected together, individually, or not at all--with varying flux densities in each case. While no confusion corrections for 60-80 mas were needed in this study, future studies should treat this range with care by determining whether \textit{Starfinder} is detecting Sgr A* and the confused star separately or together. Beyond 80 mas, Sgr A* is unambiguously detected with photometric biases consistent with zero, so no corrections are needed for stars at these distances.

\begin{figure}
\centering
\includegraphics[width=4in]{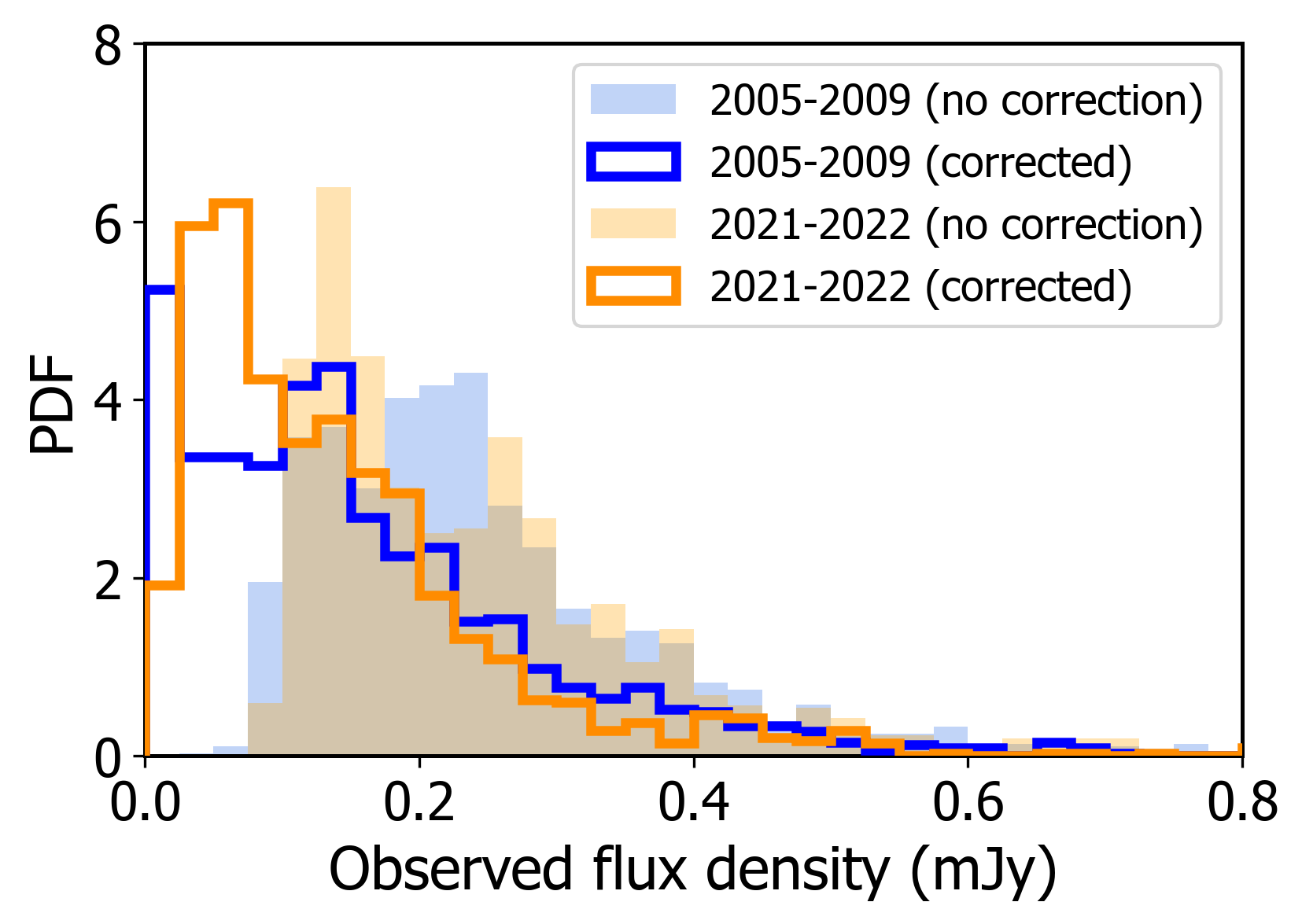}
\caption{\footnotesize Flux distributions before and after confusion corrections for epochs affected by stellar confusion. From 2005-2009 (blue), we make corrections for S0-104 and S0-102. In 2021-2022 (orange), we make corrections for S0-61 and S0-38. We see that the distributions are shifted to lower flux densities upon removal of the photometric bias. Despite reflecting different epochs that are affected by confusion from different stars, these two corrected flux distributions appear to be quite similar.}
\label{fig:confusion_histograms}
\end{figure}

For the observation epochs affected by stellar confusion, we determined the distance of the confusing star from Sgr A* by performing an orbital fit to aligned images of the Galactic center and obtaining the predicted distance at the epoch observation time (O'Neil et al. in prep). Given the brightness of the star and its distance, we interpolated the star-planting results in Table \ref{tab:starplanting} to obtain the flux bias to subtract from the data. The epoch-wise corrections we applied are given in Table \ref{tab:corrections}, and the flux distributions of affected data before and after correction are shown in Figure \ref{fig:confusion_histograms}. This procedure overcorrects and yields negative flux values in only a small percentage of frames. The corrections slightly increase the photometric uncertainties to $\sim$20\% for low fluxes, and high flux uncertainties are typically less than 5\%. In this work, we have excluded epochs in which the bright star S0-2 falls within 60 mas of Sgr A* (July 2017-2018), as the uncertainties on S0-2's flux bias correction are comparable to faint Sgr A* flux levels. \textit{Starfinder} reliably detects S0-2 and Sgr A* as distinct sources in the remaining data.

There is likely an unresolved stellar component that contributes to the measured flux densities, but we are unable to disentangle this effect from temporal variations of the Sgr A* baseline flux in this study. However, these unresolved stars are faint by nature and expected to only slightly shift the median flux density between epochs.

\begin{sidewaystable*}[ht]
\begin{center}
\setlength{\tabcolsep}{6pt}
\caption{Photometric bias measured in star-planting simulations}
\label{tab:starplanting}
\begin{tabular}{|c | c| c| c| c| c| c| c| c| c|}
\hline
\multicolumn{10}{|c|}{Photometric bias (mJy)} \\
\hline
\multicolumn{1}{|c|}{Injected star} & \multicolumn{9}{c|}{Distance planted from Sgr A* (mas)}\\
($K'$ mag) & 0 & 10 & 20 & 30 & 40 & 50 & 60-80$^a$ & 90 & 100 \\
\hline
14.0 & 1.710 $\pm$ 0.050 & 1.700 $\pm$ 0.045 & 1.678 $\pm$ 0.045 & 1.654 $\pm$ 0.038 & 1.645 $\pm$ 0.047 & 1.701 $\pm$ 0.123 & ...	& -0.002 $\pm$ 0.003 & -0.001 $\pm$ 0.007\\
\hline
14.25 & 1.348 $\pm$ 0.046 & 1.356 $\pm$ 0.057 & 1.331 $\pm$ 0.039 & 1.310 $\pm$ 0.046 & 1.278 $\pm$ 0.038 & 1.255 $\pm$ 0.027 & ... & 0.001 $\pm$ 0.007	 & -0.006 $\pm$ 0.005\\
\hline
14.5 & 1.077 $\pm$ 0.036 & 1.066 $\pm$ 0.024 & 1.028 $\pm$ 0.044 & 1.044 $\pm$ 0.029 & 1.013 $\pm$ 0.037 & 0.995 $\pm$ 0.034 & ...	& 0.001 $\pm$ 0.004 & -0.002 $\pm$ 0.004\\
\hline
14.75 & 0.856 $\pm$ 0.034 & 0.847 $\pm$ 0.031 & 0.833 $\pm$ 0.025 & 0.809 $\pm$ 0.021 & 0.769 $\pm$ 0.010 & 0.782 $\pm$ 0.041 & ...	& 	-0.007 $\pm$ 0.010 & 0.001 $\pm$ 0.006\\
\hline
15.0 & 0.671 $\pm$ 0.020 & 0.685 $\pm$ 0.039 & 0.668 $\pm$ 0.021 & 0.631 $\pm$ 0.024 & 0.649 $\pm$ 0.063 & 0.581 $\pm$ 0.030 & ...	& -0.003 $\pm$ 0.013 & -0.009 $\pm$ 0.019\\
\hline
15.25 & 0.538 $\pm$ 0.019 & 0.531 $\pm$ 0.012 & 0.506 $\pm$ 0.018 & 0.552 $\pm$ 0.082 & 0.476 $\pm$ 0.055 & 0.468 $\pm$ 0.051 & ...	& 0.006 $\pm$ 0.005 & 0.006 $\pm$ 0.004\\
\hline
15.5 & 0.425 $\pm$ 0.017 & 0.430 $\pm$ 0.013 & 0.446 $\pm$ 0.048 & 0.416 $\pm$ 0.071 & 0.371 $\pm$ 0.034 & 0.343 $\pm$ 0.016 & ...	& 0.002 $\pm$ 0.006 & -0.004 $\pm$ 0.005\\
\hline
15.75 & 0.344 $\pm$ 0.018 & 0.355 $\pm$ 0.028 & 0.337 $\pm$ 0.044 & 0.316 $\pm$ 0.071 & 0.324 $\pm$ 0.079 & 0.231 $\pm$ 0.027 & ...	& -0.006 $\pm$ 0.012 & -0.003 $\pm$ 0.012\\
\hline
16.0 & 0.278 $\pm$ 0.017 & 0.301 $\pm$ 0.019 & 0.257 $\pm$ 0.006 & 0.217 $\pm$ 0.030 & 0.200 $\pm$ 0.025 & 0.203 $\pm$ 0.002  & ... & 0.006 $\pm$ 0.004 & -0.008 $\pm$ 0.012\\
\hline
16.25 & 0.222 $\pm$ 0.019 & 0.195 $\pm$ 0.023 & 0.202 $\pm$ 0.008 & 0.196 $\pm$ 0.006 & 0.104 $\pm$ 0.012 & 0.142 $\pm$ 0.030 & ... & 0.006 $\pm$ 0.006 & -0.001 $\pm$ 0.011\\
\hline
16.5 & 0.166 $\pm$ 0.013 & 0.170 $\pm$ 0.006 & 0.165 $\pm$ 0.006 & 0.149 $\pm$ 0.010 &	0.098 $\pm$ 0.022 & 0.099 $\pm$ 0.009 & ... & -0.005 $\pm$ 0.014 & -0.009 $\pm$ 0.011\\
\hline
16.75 & 0.134 $\pm$ 0.004 & 0.132 $\pm$ 0.003 & 0.127 $\pm$ 0.006 & 0.109 $\pm$ 0.010 & 0.097 $\pm$ 0.007 & 0.057 $\pm$ 0.023 & ... & -0.006 $\pm$ 0.018 & -0.006 $\pm$ 0.004\\
\hline
17.0 & 0.101 $\pm$ 0.010 & 0.105 $\pm$ 0.003 & 0.101 $\pm$ 0.013	& 0.087 $\pm$ 0.007 & 0.070 $\pm$ 0.012 & 0.038 $\pm$ 0.009 & ... & 0.000 $\pm$ 0.006 & -0.006 $\pm$ 0.007 \\
\hline
17.25 & 0.083 $\pm$ 0.010 & 0.072 $\pm$ 0.012 & 0.057 $\pm$ 0.028 & 0.064 $\pm$ 0.013 & 0.054 $\pm$ 0.009 & 0.040 $\pm$ 0.007 & ... & -0.005 $\pm$ 0.001 & 0.001 $\pm$ 0.005\\
\hline
17.5 & 0.069 $\pm$ 0.007 & 0.066 $\pm$ 0.005 & 0.062 $\pm$ 0.006 & 0.047 $\pm$ 0.012 & 0.033 $\pm$ 0.009 & 0.003 $\pm$ 0.021 & ... & -0.001 $\pm$ 0.003 & -0.003 $\pm$ 0.005\\
\hline
17.75 & 0.054 $\pm$ 0.005 & 0.045 $\pm$ 0.009 & 0.049 $\pm$ 0.003 & 0.040 $\pm$ 0.005 & 0.036 $\pm$ 0.005 & 0.011 $\pm$ 0.010 & ... & -0.002 $\pm$ 0.007 & -0.001 $\pm$ 0.006\\
\hline
18.0  & 0.045 $\pm$ 0.003 & 0.043 $\pm$ 0.008 & 0.035 $\pm$ 0.005 & 0.025 $\pm$ 0.014 & 0.012 $\pm$ 0.008 & 0.016 $\pm$ 0.007 & ... & 0.000 $\pm$ 0.005 & -0.005 $\pm$ 0.009 \\
\hline \hline
\end{tabular}
\end{center}
$^a$ From $\sim$60-80 mas, Sgr A* and the injected star are sometimes detected together, individually, or not at all--with varying flux densities in each case. Confusion corrections in this range are unreliable. Beyond 80 mas, the photometric biases are consistent with zero, so we do not make corrections to the data in this range.
\end{sidewaystable*}

\begin{sidewaystable*}[ht]
\begin{center}
\setlength{\tabcolsep}{6pt}
\caption{Astrometric bias measured in star-planting simulations}
\label{tab:starplanting_astrometry}
\begin{tabular}{|c | c| c| c| c| c| c| c| c| c|}
\hline
\multicolumn{10}{|c|}{Astrometric bias (arcseconds)} \\
\hline
\multicolumn{1}{|c|}{Injected star} & \multicolumn{9}{c|}{Distance planted from Sgr A* (mas)}\\
($K'$ mag) & 0 & 10 & 20 & 30 & 40 & 50 & 60-80$^a$ & 90 & 100 \\
\hline
14.0 & 0.001 $\pm$ 0.001 & 0.009 $\pm$ 0.001 & 0.019 $\pm$ 0.000 & 0.029 $\pm$ 0.000 & 0.040 $\pm$ 0.002 & 0.050 $\pm$ 0.002 & ... & 0.002 $\pm$ 0.000 & 0.004 $\pm$ 0.003\\
\hline
14.25 & 0.001 $\pm$ 0.001 & 0.010 $\pm$ 0.001 & 0.019 $\pm$ 0.000 & 0.029 $\pm$ 0.001	& 0.038 $\pm$ 0.001	& 0.048 $\pm$ 0.001	& ... &	0.002 $\pm$ 0.001 & 0.005 $\pm$ 0.003\\
\hline
14.5 & 0.002 $\pm$ 0.001 & 0.010 $\pm$ 0.001 & 0.019 $\pm$ 0.000 & 0.028 $\pm$ 0.000 & 0.038 $\pm$ 0.001	& 0.048 $\pm$ 0.001 & ... & 0.004 $\pm$ 0.001 & 0.002 $\pm$ 0.001\\
\hline
14.75 & 0.002 $\pm$ 0.002 & 0.010 $\pm$ 0.001 & 0.019 $\pm$ 0.001 & 0.028 $\pm$ 0.001	 & 0.037 $\pm$ 0.001 & 0.048 $\pm$ 0.001 & ... & 0.005 $\pm$ 0.002 & 0.002 $\pm$ 0.001\\
\hline
15.0 & 0.003 $\pm$ 0.002 & 0.009 $\pm$ 0.003 & 0.019 $\pm$ 0.003 & 0.028 $\pm$ 0.001	& 0.039 $\pm$ 0.003	& 0.048 $\pm$ 0.001	& ... & 0.003 $\pm$ 0.003	& 0.005 $\pm$ 0.003\\
\hline
15.25 & 0.003 $\pm$ 0.003 & 0.011 $\pm$ 0.004 & 0.017 $\pm$ 0.001 & 0.033 $\pm$ 0.006	& 0.037 $\pm$ 0.002	& 0.047 $\pm$ 0.001	 & ... & 0.006 $\pm$ 0.004	& 0.004 $\pm$ 0.002\\
\hline
15.5 & 0.005 $\pm$ 0.003 & 0.006 $\pm$ 0.004 & 0.021 $\pm$ 0.005 & 0.027 $\pm$ 0.005	& 0.036 $\pm$ 0.012 & 0.044 $\pm$ 0.003 & ...	& 0.005 $\pm$ 0.002	& 0.003 $\pm$ 0.002\\
\hline
15.75 & 0.006 $\pm$ 0.004 & 0.013 $\pm$ 0.008 & 0.018 $\pm$ 0.004 & 0.028 $\pm$ 0.006	& 0.040 $\pm$ 0.006	& 0.047 $\pm$ 0.002	& ... & 0.005 $\pm$ 0.003 & 0.005 $\pm$ 0.003\\
\hline
16.0 & 0.008 $\pm$ 0.006 & 0.015 $\pm$ 0.004 & 0.013 $\pm$ 0.003 & 0.025 $\pm$ 0.002	& 0.036 $\pm$ 0.003	& 0.042 $\pm$ 0.001 & ... & 0.004 $\pm$ 0.002 & 0.003 $\pm$ 0.002\\
\hline
16.25 & 0.004 $\pm$ 0.005 & 0.006 $\pm$ 0.001 & 0.014 $\pm$ 0.008 & 0.022 $\pm$ 0.001	& 0.038 $\pm$ 0.001 & 0.037 $\pm$ 0.001 & ... & 0.002 $\pm$ 0.001	& 0.004 $\pm$ 0.002\\
\hline
16.5 & 0.001 $\pm$ 0.001 & 0.007 $\pm$ 0.000 & 0.013 $\pm$ 0.001 & 0.020 $\pm$ 0.001	& 0.031 $\pm$ 0.003 & 0.040 $\pm$ 0.006 & ... & 0.007 $\pm$ 0.005 & 0.004 $\pm$ 0.003\\
\hline
16.75 & 0.001 $\pm$ 0.001 & 0.006 $\pm$ 0.001 & 0.011 $\pm$ 0.002 & 0.019 $\pm$ 0.002	& 0.026 $\pm$ 0.013	& 0.031 $\pm$ 0.017	& ... & 0.004 $\pm$ 0.001	& 0.002 $\pm$ 0.001\\
\hline
17.0 & 0.002 $\pm$ 0.001 & 0.006 $\pm$ 0.001 & 0.009 $\pm$ 0.001 & 0.015 $\pm$ 0.001	& 0.021 $\pm$ 0.002	& 0.041 $\pm$ 0.006 & ... & 0.002 $\pm$ 0.002 & 0.002 $\pm$ 0.001\\
\hline
17.25 & 0.002 $\pm$ 0.001 & 0.006 $\pm$ 0.001 & 0.008 $\pm$ 0.002 & 0.014 $\pm$ 0.001	& 0.020 $\pm$ 0.001	& 0.020 $\pm$ 0.004	& ... & 0.004 $\pm$ 0.003 & 0.004 $\pm$ 0.002\\
\hline
17.5 & 0.001 $\pm$ 0.001 & 0.005 $\pm$ 0.001 & 0.008 $\pm$ 0.001 & 0.011 $\pm$ 0.003	& 0.014 $\pm$ 0.003	& 0.016 $\pm$ 0.007	& ... & 0.003 $\pm$ 0.002 & 0.003 $\pm$ 0.002\\
\hline
17.75 & 0.001 $\pm$ 0.001 & 0.006 $\pm$ 0.002 & 0.008 $\pm$ 0.001 & 0.009 $\pm$ 0.003	& 0.012 $\pm$ 0.003	& 0.008 $\pm$ 0.003	& ... & 0.005 $\pm$ 0.003 & 0.003 $\pm$ 0.002\\
\hline
18.0  & 0.001 $\pm$ 0.001 & 0.002 $\pm$ 0.001 & 0.006 $\pm$ 0.003 & 0.009 $\pm$ 0.003	& 0.005 $\pm$ 0.003	& 0.009 $\pm$ 0.002	& ... & 0.003 $\pm$ 0.002 & 0.003 $\pm$ 0.002\\
\hline \hline
\end{tabular}
\end{center}
$^a$ From $\sim$60-80 mas, Sgr A* and the injected star are sometimes detected together, individually, or not at all--with varying positions in each case. Future studies should treat this range with care by determining whether \textit{Starfinder} is detecting Sgr A* and the confused star separately or together.
\end{sidewaystable*}

\begin{table*}
\begin{center}
\setlength{\tabcolsep}{7pt}
\caption{Applied confusion corrections}
\label{tab:corrections}
\begin{tabular}{c c c c}
\hline\hline
Date (UT) & Confusing star ($K'$ mag) & Distance from Sgr A* (mas) & Confusion correction (mJy) \\
\hline
2005-07-31 & S0-104 (16.8) & 45 & 0.07 $\pm$ 0.02\\
2006-05-03 & S0-104 (16.8) & 26 & 0.11 $\pm$ 0.01\\
2006-06-20 & S0-104 (16.8) & 23 & 0.12 $\pm$ 0.01\\
2006-06-21 & S0-104 (16.8) & 23 & 0.12 $\pm$ 0.01\\
2006-07-17 & S0-104 (16.8) & 21 & 0.12 $\pm$ 0.01\\
2007-05-17 & S0-104 (16.8) & 3  & 0.13 $\pm$ 0.01\\
2007-08-10 & S0-104 (16.8) & 7  & 0.13 $\pm$ 0.01\\
2007-08-12 & S0-104 (16.8) & 7  & 0.13 $\pm$ 0.01\\
2008-05-15 & S0-104 (16.8) & 26 & 0.11 $\pm$ 0.01\\
2008-07-24 & S0-104 (16.8) & 31 & 0.10 $\pm$ 0.01\\
2009-05-01 & S0-102 (17.1) & 28 & 0.08 $\pm$ 0.02\\
2009-05-02 & S0-102 (17.1) & 28 & 0.08 $\pm$ 0.02\\
2009-05-04 & S0-102 (17.1) & 28 & 0.08 $\pm$ 0.02\\
2009-07-22 & S0-102 (17.1) & 33 & 0.07 $\pm$ 0.01\\
2009-07-24 & S0-102 (17.1) & 33 & 0.07 $\pm$ 0.01\\
2009-09-09 & S0-102 (17.1) & 38 & 0.07 $\pm$ 0.01\\
2021-05-13 & S0-61 (16.5)  & 18 & 0.17 $\pm$ 0.01\\
2021-05-14 & S0-61 (16.5)  & 18 & 0.17 $\pm$ 0.01\\
2021-07-13 & S0-61 (16.5)  & 39 & 0.10 $\pm$ 0.02\\
2021-07-14 & S0-61 (16.5)  & 39 & 0.10 $\pm$ 0.02\\
2021-08-13 & S0-61 (16.5)  & 47 & 0.10 $\pm$ 0.02\\
2022-05-14 & S0-38 (17.0)  & 26 & 0.09 $\pm$ 0.01\\
2022-05-15 & S0-38 (17.0)  & 26 & 0.09 $\pm$ 0.01\\
2022-05-21 & S0-38 (17.0)  & 26 & 0.09 $\pm$ 0.01\\
2022-05-25 & S0-38 (17.0)  & 26 & 0.09 $\pm$ 0.01\\
2022-07-19 & S0-38 (17.0)  & 33 & 0.08 $\pm$ 0.01\\
2022-07-22 & S0-38 (17.0)  & 33 & 0.08 $\pm$ 0.01\\
2022-08-14 & S0-38 (17.0)  & 38 & 0.07 $\pm$ 0.01\\
2022-08-15 & S0-38 (17.0)  & 38 & 0.07 $\pm$ 0.01\\
2022-08-16 & S0-38 (17.0)  & 38 & 0.07 $\pm$ 0.01\\
2022-08-19 & S0-38 (17.0)  & 38 & 0.07 $\pm$ 0.01\\
2022-08-20 & S0-38 (17.0)  & 38 & 0.07 $\pm$ 0.01\\
\hline \hline
\end{tabular}
\end{center}
Note: Confusion corrections estimated by interpolating results in Table \ref{tab:starplanting}. Corrections are subtracted from the $K'$ observed flux densities.
\end{table*}

\section{Structure Function Analysis}
\label{app:structurefuncs}

\begin{figure*}
\centering
\includegraphics[width=4in]{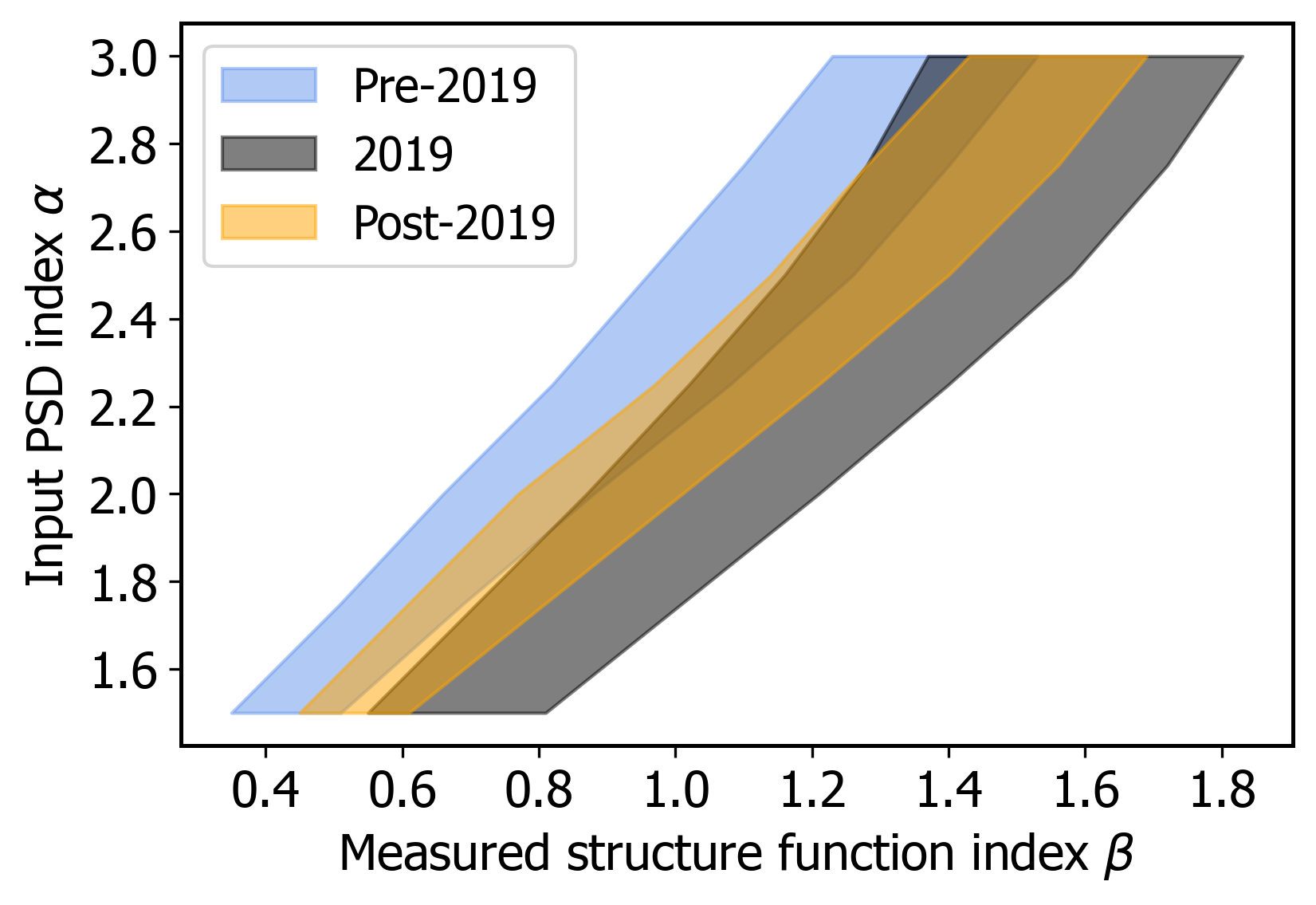}
\caption{\footnotesize 1$\sigma$ bands for the measured structure function slope $\beta$ with input PSD slope $\alpha$ for simulated light curves with the time sampling of the pre-2019 (blue), 2019 (black), and post-2019 (orange) observations. We perform a linear fit to these simulations to convert between the structure function slope and PSD slope in the data.}
\label{fig:alphabeta}
\end{figure*}

To convert between the structure function slope $\beta$ and PSD slope $\alpha$, we generate $10^3$ light curves with the respective time sampling and fixed PSD slope $\alpha$, then measure the structure function slope $\beta$. This procedure is repeated for varying values of $\alpha$. See Figure \ref{fig:alphabeta} for the relationships between $\alpha$ and $\beta$ that we find from the simulations. We perform a linear fit to these values to make the conversion between structure function slope and PSD slope in the data.

\begin{figure*}
\centering
\includegraphics[width=4in]{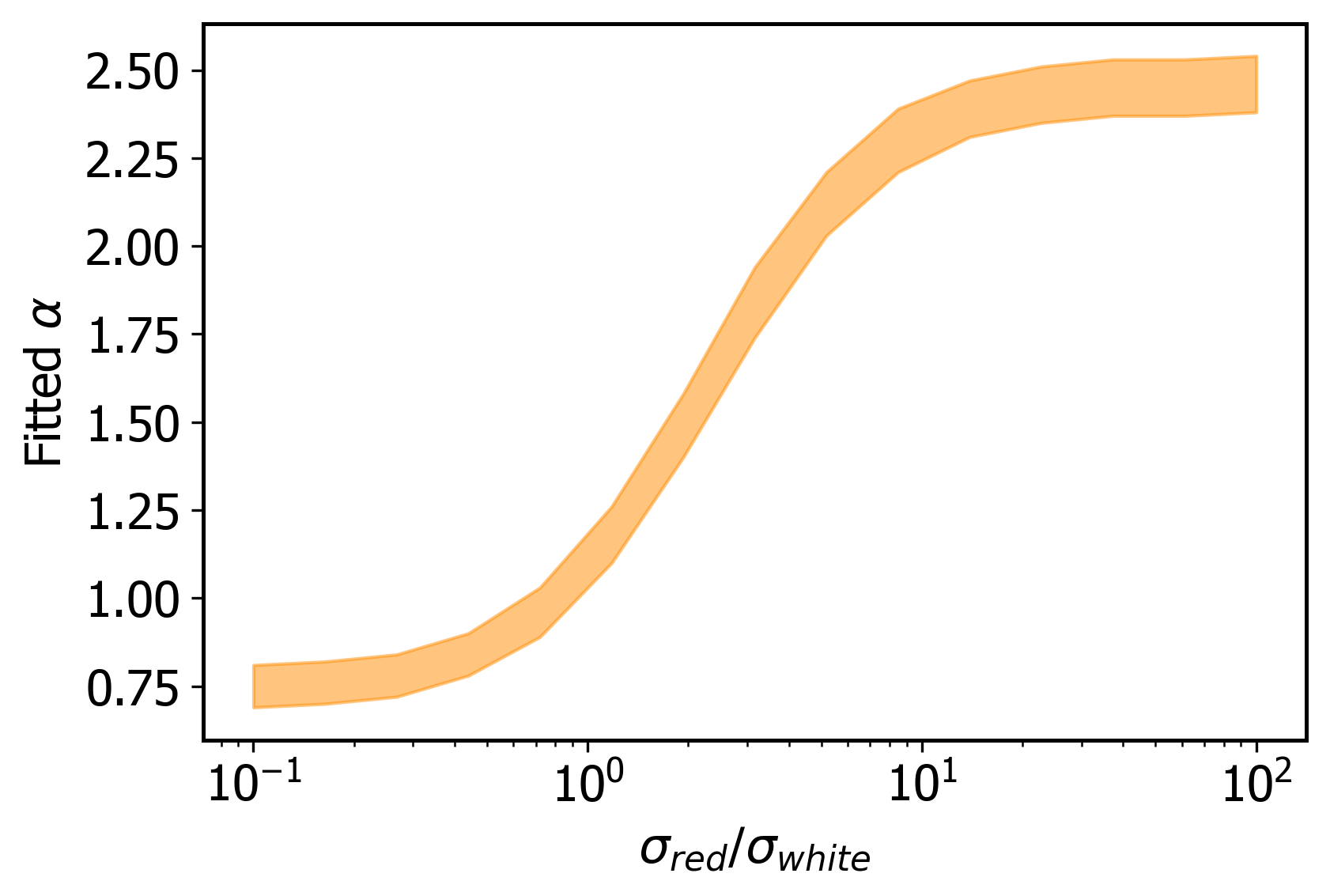}
\caption{\footnotesize 1$\sigma$ band of the fitted PSD slope $\alpha$ as a function of the ratio of the standard deviation of red noise ($\alpha = 2.5$) to the standard deviation of white noise for $10^3$ light curves with the post-2019 sampling. We see that a more prevalent white noise component causes a shallower PSD slope to be measured.}
\label{fig:rdnoisefrac}
\end{figure*}

In Table \ref{tab:PSDfits}, we found a shallower PSD slope for the 2020-2022 data compared to the pre-2019 and 2019 data. To determine whether a greater white noise component in the post-2019 data is responsible for this shallower slope, we simulate $10^3$ light curves with the post-2019 time sampling assuming an intrisic PSD slope of $\alpha = 2.5$, then add varying levels of white noise and fit for a measured value of $\alpha$ \citep[a procedure performed in][]{Do19}. We find in Figure \ref{fig:rdnoisefrac} that as the ratio of white noise to red noise becomes $\lesssim$10, a flatter PSD slope is indeed measured. As a result, the white noise contamination from the stars S0-61 in 2021 and S0-38 in 2022 could have caused a shallower value of $\alpha$ to be measured for these years.

\end{document}